%Character Formulae and Partition Functions in 
%$d$-Dimensional Conformal Field Theory
\input harvmac
%\draft
\input amssym.def
\input amssym
\baselineskip 14pt
\magnification\magstep1
\font\bigcmsy=cmsy10 scaled 1500
\parskip 6pt
%For smaller font footnotes
\newdimen\itemindent \itemindent=32pt
\def\textindent#1{\parindent=\itemindent\let\par=\resetpar%
\indent\llap{#1\enspace}\ignorespaces}

\let\oldpar=\par
\def\resetpar{\oldpar\parindent=20pt\let\par=\oldpar}

\font\ninerm=cmr9 \font\ninesy=cmsy9
\font\eightrm=cmr8 \font\sixrm=cmr6
\font\eighti=cmmi8 \font\sixi=cmmi6
\font\eightsy=cmsy8 \font\sixsy=cmsy6
\font\eightbf=cmbx8 \font\sixbf=cmbx6
\font\eightit=cmti8
\def\eightpoint{\def\rm{\fam0\eightrm}
  \textfont0=\eightrm \scriptfont0=\sixrm \scriptscriptfont0=\fiverm
  \textfont1=\eighti  \scriptfont1=\sixi  \scriptscriptfont1=\fivei
  \textfont2=\eightsy \scriptfont2=\sixsy \scriptscriptfont2=\fivesy
  \textfont3=\tenex   \scriptfont3=\tenex \scriptscriptfont3=\tenex
  \textfont\itfam=\eightit  \def\it{\fam\itfam\eightit}%
  \textfont\bffam=\eightbf  \scriptfont\bffam=\sixbf
  \scriptscriptfont\bffam=\fivebf  \def\bf{\fam\bffam\eightbf}%
  \normalbaselineskip=9pt
  \setbox\strutbox=\hbox{\vrule height7pt depth2pt width0pt}%
  \let\big=\eightbig  \normalbaselines\rm}
\catcode`@=11 %
\def\eightbig#1{{\hbox{$\textfont0=\ninerm\textfont2=\ninesy
  \left#1\vbox to6.5pt{}\right.\n@@space$}}}
\def\vfootnote#1{\insert\footins\bgroup\eightpoint
  \interlinepenalty=\interfootnotelinepenalty
  \splittopskip=\ht\strutbox %
  \splitmaxdepth=\dp\strutbox %
  \leftskip=0pt \rightskip=0pt \spaceskip=0pt \xspaceskip=0pt
  \textindent{#1}\footstrut\futurelet\next\fo@t}
\catcode`@=12 %

\def \de{\delta}
\def \De{\Delta}
\def \si{\sigma}

\def \al{\alpha}
\def \be{\beta}
\def \pr{\partial}

\def \l{\big \langle}
\def \r{\big \rangle}
\def \ep{\epsilon}
\def \vep{\varepsilon}
\def \half{{\textstyle {1 \over 2}}}

\def \ts{\textstyle}
\def \A{{\cal A}}
\def \B{{\cal B}}
\def \C{{\cal C}}
\def \D{{\cal D}}

\def \H{{\cal H}}
\def \I{{\cal I}}

\def \K{{\cal K}}

\def \S{{\cal S}}

\def \W{{\cal W}}
\def \V{{\cal V}}

\def \e{{\rm e}}
\def \x{{\rm x}}

\def \bj{\bar \jmath}
\def \al{\alpha}

\def \e{{\rm e}}
\def \uell{\underline \ell}
\def \ue{\underline{\rm e}}
\def \rhw{{\rm h.w.}}
\def \v{{\rm v}}
\def \P{{\cal P}}
\def \uv{{\underline{\rm v}}}
\def \x{{{\rm x}}}
\font \bigbf=cmbx10 scaled \magstep1
\def \ulambda{{\underline{\Lambda}}}
\def \mapright#1{\smash{\mathop{\longrightarrow}\limits^{#1}}}
\def \mapup#1{\smash{\mathop{\nearrow}\limits^{#1}}}
\def \mapdown#1{\smash{\mathop{\searrow}\limits_{#1}}}
\def \mapupo#1{\smash{\mathop{\nearrow}\limits_{#1}}}
\def \mapdowno#1{\smash{\mathop{\searrow}\limits^{#1}}}
\def \sm{\smash}
\def \Bsw{\!\mathrel{\hbox{\bigcmsy\char'056}}\!}
\def \Bse{\!\mathrel{\hbox{\bigcmsy\char'046}}\!}

\def \Geq{\geqslant}
\def \leq{\leqslant}
 \def \geq{\geqslant}
\def \T{{\cal T}}

%References
\lref\Bianchi{M. Bianchi, J.F. Morales and H. Samtleben,
On stringy AdS$_5\times S^5$ and higher spin 
holography, JHEP 0307 062 2003, hep-th/0310129.}
\lref\Bianchit{N. Beisert, M. Bianchi, J.F. Morales and H. Samtleben,
On the spectrum of AdS/CFT beyond supergravity, JHEP 0402 001 2004,
hep-th/0310292.}
\lref\Bara{A. Barabanschikov, L. Grant, L.L. Huang and S. Raju,
The spectrum of Yang Mills on a sphere, hep-th/0501063.}
\lref\Flato{M. Flato and C. Fronsdal, One massless particle equals
two Dirac singletons, Lett. Math. Phys. 2 421 1978.}
\lref\dobo{ V.K. Dobrev and E. Sezgin,
Spectrum and Character Formulae of $so(3,2)$ Unitary Representations,
Lecture Notes in Physics,
Vol. 379, Springer-Verlag, Berlin, 1990, pp. 227-238;
eds. J.D. Hennig, W. L\"{u}cke and J. Tolar.}

\lref\dobtw{V.K. Dobrev,
Positive energy representations of noncompact quantum algebras,
Proceedings of the Workshop on Generalized
Symmetries in Physics,
Clausthal, July 1993, World Sci. Singapore, 1994, pp. 90-110;  eds.
H.D. Doebner et al.
}
\lref\angel{E. Angelopoulos and M. Laoues, Singletons on ADS$_n$,
Dijon 1999, Quantization, deformations and symmetries, vol. 2 3-23.}
\lref\Vas{M.A. Vasiliev, Higher spin superalgebras in any dimension
and their representations, JHEP 0412 0426 2004, hep-th/0404124.}
\lref\gun{M. G\"unaydin and N. Marcus, The spectrum of the $S^{5}$
compactification of the chiral $N=2$, $D=10$ supergravity
and the unitary supermultiplets of $U(2,2|4)$, Class. Quant. Grav. 2 L11 1985.}
\lref\sez{E. Sezgin and P. Sundell, Massless higher spins and
holography, Nucl. Phys. B 634 120 2002, hep-th/0112100.}
\lref\kut{D. Kutasov and F. Larsen, Partition sums and entropy
bounds in weakly coupled CFT, JHEP 0101 001 2001, hep-th/0009244.}
\lref\aha{O. Aharony, J. Marsano, S. Minwalla, K. Papadodimas and
M. Van Raamsdonk, The Hagedorn deconfinement phase transition in
weakly
coupled large $N$ gauge theories, Varna 2003, Lie theory and its
applications in physics V, 161-203, hep-th/0310285.}
\lref\gib{G.W. Gibbons, M. Perry and C.N. Pope, Bulk and boundary partition
functions for AdS, in preparation.}
\lref\cardy{J.L. Cardy, Operator content and modular properties
of higher dimensional conformal field theories, Nucl. Phys. B366 403 1991.}
\lref\ferr{S. Ferrara and C. Fronsdal, Conformal fields in higher
dimensions,
Rome 2000, Recent developments in theoretical and experimental
general relativity, gravitation and relativistic 
field theories, Pt. A, 508-527, hep-th/0006009.}
\lref\min{S. Minwalla, Restrictions imposed by superconformal
invariance on Quantum field theories, Adv. Theor. Math. Phys. 2 781
1998,
hep-th/9712074.}
\lref\fult{W. Fulton and J. Harris,
Representation theory, a first course, Graduate texts in
mathematics, Springer-Verlag New York, 1991.
} 
\lref\sie{W. Siegel, All free conformal representations in
all dimensions, Int. J. Mod. Phys. A4 2015 1989.}
\lref\mack{G. Mack, All unitary ray representations of the conformal
group $SU(2,2)$ with positive energy, Commun. Math. Phys. 55 1 1977.}
\lref\Dobrev{V.K. Dobrev, Characters of the positive energy UIRs of 
$D=4$ conformal supersymmetry, hep-th/0406154.}
\lref\gru{B. Gruber and A.U. Klimyk, Properties of linear
representations with a highest weight for the semi-simple 
Lie algebras, J. Math. Phys. Vol. 16 (1975) 1816.}
\lref\ber{I.N. Bernstein, I.M. Gel'fand and S.I. Gel'fand,
Structure of representations generated by vectors of highest weight, Funct.
Annal. App. 5 (1971) 1.}
\lref\ver{Daya-Nand Verma,
Structure of certain induced representations 
of complex semisimple Lie algebras, Bull. Am. Math. Soc. 74 (1968) 160.}
\lref\fuchs{J. Fuchs and C. Schweigert,
Symmetries, Lie algebras and representations, CUP, 1997.}
\lref\hum{J.E. Humphreys, Reflection groups and Coxeter
groups, CUP, 1990.}
\lref\enr{T. Enright, Analogues of Kostant's $u$-cohomology
formulas for unitary highest weight modules, J. Reine Angew. Math.,
392 (1988) 27.}

{\nopagenumbers
\rightline{DAMTP/05-69}
\rightline{hep-th/0508031}
\vskip 1.5truecm
\centerline {\bigbf Character Formulae and Partition Functions}
\vskip 4pt
\centerline {\bigbf in Higher Dimensional Conformal Field Theory}
\vskip  6pt
%\centerline
\vskip 2.0 true cm
\centerline {F.A. Dolan ${}^\dagger$}

\vskip 12pt
\centerline {\ Department of Applied Mathematics and Theoretical Physics,}
\centerline {Wilberforce Road, Cambridge CB3 0WA, England}
%\centerline {\ Department of Applied Mathematics and Theoretical Physics,}
%\centerline {Silver Street, Cambridge, CB3 9EW, England}
\vskip 1.5 true cm

{\eightpoint
\parindent 1.5cm{
\noindent
A discussion of character formulae for
positive energy
unitary irreducible representations of the the conformal group 
is given, employing Verma
modules and Weyl group reflections.  
Product formulae for
various conformal group representations are 
found.  These include generalisations
of those found by Flato and Fronsdal for $SO(3,2)$.
In even dimensions the products
for free representations split
into two types depending on whether the
dimension is divisible by four or not.
{\narrower\smallskip\parindent 0pt


\narrower}}

\vfill
\line{${}^\dagger$ 
address for correspondence: Trinity College, Cambridge, CB2 1TQ, England\hfill}
\line{\hskip0.2cm email:
{{\tt fad20@damtp.cam.ac.uk}}\hfill}
}
%\vskip0.5cm

%PACS: 11.10-z; 11.25.Hf; 11.10Kk; 04.62+v
\eject}
\pageno=1
\newsec{Introduction}
Motivated by the AdS/CFT correspondence, character formulae
for groups associated with conformal symmetry have received greater
attention recently \refs{\Bianchi,\Bianchit,\Dobrev, \Bara}.
In \refs{\Bianchi,\Bianchit,\Bara} the way in which 
character formulae
encode the spectrum of operators allowed in a
conformal Yang-Mills theory 
has been their main use.  We hope that the
present discussion might be similarly useful for
conformal Yang-Mills theories in higher
dimensions.

It is well known that character formulae 
provide an elegant way of decomposing tensor products 
of Lie algebra representations - the Racah-Speiser 
algorithm for decomposing tensor products
of finite dimensional irreducible representations of simple
Lie algebras  may be
easily proved in terms of Weyl characters,
see \fuchs\ for a summary.
In conformal field theories the method
of characters was used by Flato and Fronsdal \Flato\
to decompose products of certain massless
representations, called `Di' and `Rac', in three dimensions.
Oscillator and other methods have been
used by various authors to generalise
 this to higher dimensions
\refs{\gun, \sez, \angel, \Vas}.  Here
we follow a more direct approach using character formulae 
for the conformal group to decompose products
of positive energy unitary irreducible representations
of the conformal group which {\it inter alia} provides a generalisation
of the Flato-Fronsdal results.
These formulae may also
be relevant to operator
product expansions.     

The layout of the paper is as follows.  We 
rewrite the conformal algebra in terms of
the orthonormal basis of $SO^*(d+2)$,
the complexification of the conformal
group in $d$ dimensions, in section 2.  

In section 3 we construct the characters of
any positive energy unitary irreducible representation of
the conformal group.    The problem
is related to finding characters of certain infinite
dimensional representations of $SO^*(d+2)$
and we make use of a result in \gru,
employing Verma module characters, for
solving it.
The main part of the task consists,
in this approach, of finding sub-Verma modules
of an original one.
This is more straightforward in the
orthonormal basis of $SO^*(d+2)$
due to simplifications in the Weyl group action
on weights in this basis.
In this section we also show how these formulae
are equivalent to ones obtained as follows.
The basis for the original $SO^*(d+2)$ Verma module
is reduced in a way determined by appendix C
which discusses unitary representations of the 
conformal group.
 We write
down the character of the reduced 
$SO^*(d+2)$ Verma
module and then simply act on it with the
Weyl symmetry operator of $SO(d)$.
The formula obtained
agrees with the character formula for
the corresponding irreducible
representation of $SO^*(d+2)$.
  We also give the three and
four dimensional results explicitly
and these match known results \refs{\Bara, \Flato,\dobo,\dobtw}.

In section 4 we discuss products of the
unitary irreducible representations.
As a simple example we first discuss
the case of $d=4$.  This is made simpler
by the fact that the $SO(4)$ character
may be rewritten as a product of two $SO(3)$
characters.  We then go on to
discuss higher dimensional cases
which correspond to products of free
representations.  
Crucial in this approach are expansion
formulae of the characters in the following
form, namely,
\eqn\formcruc{
\sum_{N}s^{\De+N}F_N(\x)\,,
}
where $\De$ denotes the canonical conformal
dimension, $F_N(\x)$ is some linear combination
of the $SO(d)$ characters and with $s$ and ${\underline
\x}=(x_1,\dots,x_r),\,r=[{1\over 2}d]$
being some variables.  We give 
expansion formulae of the type \formcruc\
for all character formulae of interest.

In even dimensions the product formulae
obtained divide into two forms depending
on whether the dimension $d$ is divisible by
four or not.

While we do not find all such product
formulae, we feel that
the method presented generalises easily
when used in conjunction with expansion formulae
of the type \formcruc.

Expansions of the form \formcruc\ are 
used in section 5 then to correlate our results for character
formulae with one-particle partition functions which
have been found by various
authors \refs{\cardy, \kut, \aha}
for the free scalar, Weyl fermion and 
${d\over 2}$-form field strengths.
We also discuss an expansion formula
given for the character for
conserved symmetric traceless tensor currents in the main text.
A simple argument is given which explains the behaviour
of the character formula (in the form \formcruc)
when ${\underline \x}=(1,\dots,1)$.
Also it is found that 
the character for the free scalar obtained
here matches the one particle partition function
for a scalar field on the boundary of
AdS in \gib\ when the spin of descendants is taken into account.

Various useful formulae and proofs are 
left in the remaining appendices.
In appendix A results for
character formulae for infinite
dimensional representations of semi-simple Lie groups
are discussed.  In appendix B standard formulae
for $SO(d)$ Weyl characters are given. 
  In appendix C unitarity
bounds are discussed.
While unitarity bounds have been
discussed in great detail by
other authors,
see for example \refs{\ferr,\mack,\min},
we feel that this attention is merited
in that it determines which combination of
generators are to be omitted
from the full Verma module for $SO^*(d+2)$
when character formulae
for unitary irreducible representations
of the conformal group are analysed.
Appendix D contains
proofs of certain
product and expansion formulae for conformal
group characters.

\newsec{The conformal algebra in the orthonormal basis}

Starting from the Lie algebra for $SO(d,2)$,
\eqn\coonejk{
[M_{AB},M_{CD}] =i(g_{AC}\,M_{BD}-g_{AD}\,M_{BD}-g_{BC}\,M_{AD}+
g_{BD}\,M_{AC})\,,
}
for $A,B=1,\dots,d{+2}$,
$g_{AB}={\rm diag.}(1,\dots,1,-1,-1)$ and where
$M_{AB}=-M_{BA}$ are Hermitian, then
\coonejk\ may be related to the standard form
of the conformal algebra by defining
$M_{ab},\,\P_a,\,\K_a,\,H,$ for $a,b=1,\dots,d$ through
\eqn\defjr{
[M_{AB}]=\pmatrix{M_{ab} & -M_{at}\cr
M_{sb} & \ep_{st}H}\,,
}
where $s,t=d+1,d+2$, $\ep_{d\,d{+1}}=1,\,\ep_{st}=-\ep_{ts}$ and 
\eqn\defkl{
\P_a=M_{a\,d+1}+i M_{a\,d+2}\,,\qquad \K_a=M_{a\,d+1}-i M_{a\,d+2}\,.
}  
Then
\eqn\coone{\eqalign{
[M_{ab},M_{cd}]& {} =i(\de_{ac}\,M_{bd}-\de_{ad}\,M_{bc}-\de_{bc}\,M_{ad}+
\de_{bd}\,M_{ac})\,,\cr
[M_{ab},\P_c]& {} =i(\de_{ac}\P_b-\de_{bc}\P_a)\,,\quad
[M_{ab},\K_c]=i(\de_{ac}\K_b-\de_{bc}\K_a)\,, \cr
[H,\P_{a}] & {} =\P_{a}\,,\quad [H,\K_{a}]=-\K_{a}\,,\quad [\K_a,\P_b]=-2i
M_{ab}+2 \de_{ab}H\,, }
}
with
all other commutators not mentioned in \coone\ vanishing.
As usual, $M_{ab}=-M_{ba}$ are Hermitian generators of $SO(d)$ rotations.
Here $H$, the conformal Hamiltonian, is required to have a
positive definite spectrum of eigenvalues for positive energy
representations.

In \coone\ $M_{ab}$ of course satisfy the Lie algebra of $SO(d)$. 
For what follows we will use the orthonormal basis
for $SO(d)$ whereby the Cartan subalgebra is defined by
\eqn\cotwo{
H_i=M_{2i-1\,2i}\,,\qquad i=1\,,\dots,r\,,\qquad [H_i,H_j]=0\,,
}
with raising and lowering operators formed
from
\eqn\cothree{
E_{\sm{ij}}^{\vep\eta}=-E_{ji}^{\eta\vep}=M_{2i-1\,2j-1}+i\vep M_{2i\,2j-1}
+i\eta M_{2i-1\,2j}-\vep\eta M_{2i\,2j}\,,\qquad i\neq j\,,\quad
\vep,\eta=\pm \,,
}
augmented by
\eqn\cofour{
E_{\sm i}^{\pm}=M_{2i-1\,2r+1}\pm i M_{2i\,2r+1}\,,
}
for $SO(2r+1)$. 

In $2r$ dimensions the commutation relations among the
$SO(2r)$ generators in the orthonormal basis are given by
\eqn\coeight{\eqalign{
&[H_i,E_{jk}^{\vep\eta}]=\big(\vep\de_{ij}+\eta \de_{ik}\big)
E_{jk}^{\vep\eta}\,,\cr
& [E_{ij}^{\vep\eta},E_{ij}^{\vep'\eta'}] =(\vep-\vep')(1-\eta\eta')H_i
+(\eta-\eta')(1-\vep\vep')H_j\,,\cr
& [E_{ij}^{\vep\eta},E_{jk}^{\vep'\eta'}] =i(\vep'\eta-1)
E_{ik}^{\vep\eta'}\,,\quad i\neq k\,,\cr}
}
where $i,j,k=1,\dots,r$ and  $\vep,\vep',\eta,\eta'=\pm$
with other commutators, that can not be obtained through
the symmetry $E_{ij}^{\vep\eta}=-E_{ji}^{\eta\vep}$, vanishing.

Using standard orthogonal unit vectors $\e_i\in\Bbb{R}^r$
then $E^{+\pm}_{ij}$ for $1\leq i<j\leq r$
 correspond to the set of positive roots
$\e_i\pm \e_j$ while $E^{-\pm}_{ij}$
correspond to the set of negative roots
$-\e_i\pm \e_j$.
The simple roots 
$\e_i-\e_{i+1},\,\e_{r-1}+\e_r$
for $1\leq i\leq r{-1}$ correspond to
the linearly independent set of raising operators
$E_{i\,i+1}^{+-}$, $E_{r-1\,r}^{++}$. 
Similarly the linearly independent set of
lowering operators is
$E_{i\,i+1}^{-+},\,1\leq i\leq r-1$, $E_{r-1\,r}^{--}$.

In $2r+1$ dimensions we have additionally the following
commutation relations involving the extra generators \cofour,
namely,
\eqn\coeightoo{\eqalign{
& [H_i,E_{j}^{\pm}]=\pm \de_{ij}\,E_{j}^{\pm}\,,\cr
& [E_{i}^{\vep},E_{i}^{\eta}]=(\vep-\eta)H_i\,,\qquad 
[E_{i}^{\vep},E_{j}^{\eta}]=iE_{ij}^{\vep\eta}\,,\quad i\neq j\,,\cr
& [E_{ij}^{\vep\eta},E_{j}^{\vep'}]=-[E_{ji}^{\eta\vep},E_{j}^{\vep'}]=
i(\vep'\eta-1)E_{i}^{\vep}\,,}
}
for $i,j=1,\dots, r$ with all other such vanishing.

For $SO(2r+1)$
$E_{i}^{+}$ corresponds to the extra positive roots
$\e_i$ while $E^{-}_i$ corresponds to the extra negative
roots $-\e_i$.
The simple roots $\e_i-\e_{i+1},\,1\leq
i\leq r{-1}$ and $\e_r$  
correspond to the linearly independent set of 
raising operators $
E_{i\,i+1}^{+-}$, $E_{r}^{+}$.
The linearly
independent set of lowering operators is
$E_{i\,i+1}^{-+},\,1\leq i\leq r-1$ along with $E_{r}^{-}$.

In the orthonormal basis
of $SO(2r)$, the $2r$ dimensional 
vector representation has highest
weight $\e_1$ and all other weights in the weight system
are given by $\pm \e_i,\,1\leq i\leq r$.
The weight system may be
 represented diagrammatically by
\eqn\weightsystemeven{
\e_1
\mapright{E_{12}^{-+}}\e_2\cdots\e_{r-2}\mapright{E_{r-2\,r-1}^{-+}}
\e_{r-1}
\matrix{
& \e_r & \cr
&&\cr
\!\!\!\!\!\!\mapup{E_{r-1\,r}^{-+}}\!\!\!\!\!\! & 
&\!\!\!\!\!\! \mapdowno{E_{r-1\,r}^{--}}\cr
\!\!\!\!\!\!\mapdown{E_{r-1\,r}^{--}}\!\!\!\!\!\! & 
&\!\!\!\!\!\! \mapupo{E_{r-1\,r}^{-+}} \cr
&&\cr
& -\e_r & }
\!\!\!\!\!\!-\e_{r-1}\mapright{E_{r-2\,r-1}^{-+}}-\e_{r-2}\cdots
-\e_2\mapright{E_{1\,2}^{-+}}-\e_1
}
which also indicates the action of the lowering operators
$E_{i\,i+1}^{-+},\,1\leq i\leq r$, $E_{r-1\,r}^{--}$.
As a convenient basis for the $\P_a,\,\K_a$
operators, which of course transform in the vector
representation, we define 
\eqn\conine{
\P_{i\pm}=\P_{2i-1}\pm i  
\P_{2i}\,,\quad \K_{i\pm}=\K_{2i-1}\pm i\K_{2i}\,,
}
for which $[E_{i\,i+1}^{+-},\P_{1+}]=[E_{r-1\,r}^{++},\P_{1+}]=
[E_{i\,i+1}^{+-},\K_{1+}]=[E_{r-1\,r}^{++},\K_{1+}]=0$ and
\eqn\coten{
[H_i,\P_{j\pm}]  =\pm \de_{ij} \,\P_{j\pm}\,,\qquad
[H_i,\K_{j\pm}]  =\pm \de_{ij} \, \K_{j\pm}\,,
}
so that $\P_{i\pm},\,\K_{i\pm}$ correspond to the weights $\pm \e_i$.
In terms of explicit action of lowering operators,
we have that
\eqn\coeleven{\eqalign{
\P_{i\pm} & {}=-{\ts{i\over 2}}[E_{1i}^{-\pm},\P_{1+}]\,,\quad i=2,\dots ,
r-1\,,\quad 
\P_{1-}={\ts{1\over 4}}[E_{12}^{-+},[E_{12}^{--},\P_{1+}]]\cr
\K_{i\pm}& {} =-{\ts{i\over 2}}[E_{1i}^{-\pm},\K_{1+}]\,,\quad i=2,\dots ,
r-1\,,\quad 
\K_{1-}={\ts{1\over 4}}[E_{12}^{-+},[E_{12}^{--},\K_{1+}]]
\,,}
}
which may be easily unravelled in terms of the action of the
linearly independent set of lowering operators 
$E_{i\,i+1}^{-+},\,1\leq i\leq r$, $E_{r-1\,r}^{--}$.

In fact the entire conformal algebra in this
basis my be written in terms of the orthonormal
basis of the $SO^*(2r+2)$ algebra.
Making the definitions
\eqn\one{ 
H_0=-H\,,\quad
E_{0i}^{-\pm}=-E_{i0}^{\pm-}=\P_{i\pm}\,,
\quad 
E_{i0}^{\pm+}=-E_{0i}^{+\pm}=\K_{i\pm}\,,\quad i=1,\dots, r\,,
}
(so that $-H$ is the extra Cartan subalgebra element and
$\K_{i\pm}/ \P_{i\pm}$ form the extra raising/lowering 
operators) then the conformal algebra may be shown to be
equivalent to \coeight\ for the range of the
indices $i,j,k$ being extended to $0,1,\dots,r$.
The linearly
independent set of raising/lowering operators in this case is extended
to
include $\K_{1-}$/$\P_{1+}$.

The $2r{+1}$ dimensional 
vector representation of $SO(2r+1)$
has highest weight $\e_1$
and the weight system is  $\pm \e_i$ and $0$.
Diagrammatically the weight system  is given by
\eqn\weightsystemodd{
\e_1 \mapright{E_{12}^{-+}}\e_2\cdots\e_{r-1}\mapright{E_{r-1\,r}^{-+}}\e_{r}
\mapright{E_{r}^{-}}\,0\,\mapright{E_{r}^{-}}-\e_r
\mapright{E_{r-1\,r}^{-+}}-\e_{r-1}\cdots
-\e_2\mapright{E_{1\,2}^{-+}}-\e_1
}
where we have indicated the action of lowering operators.  
In addition to  
\conine\ we may also choose
\eqn\cofrumpet{
\P_0=\P_{2r+1}\,,\qquad \K_0=\K_{2r+1}\,,
}
as extra elements of the basis for $\P_a,\,\K_a$ operators
and these commute with $H_i,\,1\leq i\leq r$, so that they have weight
$0$.

Again, in this basis
the conformal algebra may be written in terms of the orthonormal
basis of $SO^*(2r+3)$. 
In addition to
\one\ in this case we define
\eqn\extrad{
E_{0}^{-}=\P_{0}\,,\qquad -E_{0}^{+}=\K_{0}\,,
}
and along with \coeight\ the extra commutation relations
are given by \coeightoo\ for the range of the
indices $i,j$ being extended to $0,1,\dots,r$.
Again, the linearly
independent set of raising/lowering operators in this case is extended
to
include $\K_{1-}$/$\P_{1+}$.

In terms of the orthonormal basis,
unitarity requires that
\eqn\unitarityrest{
H{}^{\dagger}=H\,,\quad H_i{}^{\dagger}=H_i\,,
\quad E_{ij}^{\vep\eta}{}^{\dagger}=E_{ij}^{-\vep\,-\eta}\,,
\quad E_{i}^{\vep}{}^{\dagger}=E_{i}^{-\vep}\,,\quad \P_{i\vep}{}^{\dagger}
=\K_{i\,-\vep}\,,\quad \P_{0}{}^{\dagger}=\K_0\,.
}

\newsec{Character formulae for positive 
energy unitary irreducible representations}

Essential in our approach to finding the character formulae
for positive energy irreducible representations of the
conformal group are $SO^*(d+2)$ Verma modules.
These have basis generated by the
arbitrary action of  
lowering operators on the conformal highest weight state
$|\De,\uell\r^{\rhw}$ corresponding to the
$SO^*(d+2)$ weight $\ulambda=(-\De,\ell_1,\dots,\ell_r),\,r=[\half d]$
(where for $a=0,1,\dots,r$ then
$\Lambda_a$ are $H_a=-H,H_i$ eigenvalues).
In what follows we assume that $\ell$ is a
dominant integral highest weight with respect
to $SO(d)$, so that in the orthonormal basis $\ell_i=\ell\cdot \e_i\in
\half\Bbb{Z}$ and{\foot{The Dynkin labels for even $d$ are
given by $\Lambda'{}_i=\ell_i-\ell_{i+1},\,1\leq i\leq r{-2}$
and $\Lambda'{}_{r-1}=\ell_{r-1}+\ell_r,\,\Lambda'{}_r=\ell_{r-1}-\ell_r$
while for odd $d$ they are  
$\Lambda'{}_i=\ell_i-\ell_{i+1},\,1\leq i\leq r{-1}$
and $\Lambda'{}_r=2\ell_r$ and these are the conditions for
them to be non-negative integers.
}}
\eqn\piffle{
\ell_1\geq\dots \geq\ell_{r-1}\geq
|\ell_r|\,,}
for $SO(2r)$
while for $SO(2r+1)$ 
\eqn\piffles{
\ell_1\geq\dots \geq\ell_r\geq 0\,,}
these being the respective dominant Weyl chambers
(or boundaries thereof).

The highest weight state $|\De,\uell\r^{\rhw}$ satisfies
\eqn\defhigh{
H_{a}|\De,\uell\r^{\rhw}=\Lambda_a |\De,\uell\r^{\rhw}\,,\qquad
\K_{1-}|\De,\uell\r^{\rhw}=E_{\al}|\De,\uell\r^{\rhw}=0\,,
}
for $\al$ being the simple roots of $SO(d)$ so that
\eqn\annih{
\{E_\al\}\to \{E^{++}_{i\,i{+1}},\,1\leq i\leq
r{-1}\,,\,\,E^{++}_{r{-1}\,r}\}\,,
}
for $SO(2r)$ while
\eqn\anniho{
\{E_\al\}\to \{E^{++}_{i\,i{+1}},\,1\leq i\leq
r{-1}\,,\,\,E^{+}_{r}\}\,,
}
for $SO(2r+1)$.

The Verma module $\V_{\Lambda}$
with highest weight $\Lambda$ therefore has basis
\eqn\cotwentythree{
\prod_{{v=i\vep,0\atop 1\leq i\leq r,\,\vep=\pm}}
\P_{v}{}^{n_{v}}\,
\prod_{\al\in\Phi_-}E_{\al}{}^{n_{\al}}|\De,\uell\r^{\rhw}\,,
}
for $\Phi_-$ denoting the set of negative 
roots of $SO(d)$
and with $n_{v},n_{\al}$ all
positive or zero integers, with
$n_0=0$ for $SO(2r)$.
As mentioned before, for $SO(2r)$
then $\{E_{\al}\}\to \{E_{ij}^{-\pm}\}$ while
for $SO(2r+1)$ then $\{E_{\al}\}\to \{E_{ij}^{-\pm},\,E_{i}^{-}\}$.
Corresponding to the basis \cotwentythree\
the weights $\Lambda'$ in the Verma module are given by
\eqn\weights{
\Lambda'_0=-\De-\!\!\!\sum_{{v=i\vep,0\atop 1\leq i\leq r,\,\vep=\pm}}
\!\!\!n_v\,,\qquad \ell'=\ell-\sum_{\al\in\Phi_-}n_{\al}
\al
+\sum_{i=1}^r \big(n_{i+}-n_{i-}\big)\e_i\,.
}
In \cotwentythree\ we assume
some fixed ordering of $\P_v,\,E_{\al}$. This
ordering may be arbitrarily chosen
since if a different ordering is assumed
then the resulting Verma module basis can be expressed in terms
of that in \cotwentythree\ due to  $\P_v,\,E_{\al}$
having commutators which are closed among themselves.

In appendix C we will use the form of the algebra 
in the last section,
in terms of the orthonormal basis, to derive conditions
necessary for conformal
group representations to be unitary.
These results are summarised by
\eqn\unitarityd{\eqalign{
\De &{} \geq \De_p=\ell_1+d-p-1\,,\quad p=1,\dots,[\half d]\quad{\rm
for}\quad 
\ell_1=\ell_2=\dots =\ell_p>\ell_{p+1}
\,,\cr
\De &{} \geq \half d-1\quad {\rm or}\quad \De=0\quad {\rm for}\quad
\uell=0\,,}
}
while for odd $d$ we have in addition that
\eqn\unitarityodd{
\De\geq [\half d]\quad {\rm for}\quad \ell_1=\dots =\ell_{[{1\over 2}
d]}
=\half\,.
}
It has been proven elsewhere that these conditions are
sufficient \ferr\ (in order for states in irreducible
representations of the conformal group to
have strictly positive norm).
We impose these conditions on the representations
we are interested in.

Along with the highest weight $\Lambda$ the
weight system for $\V_\Lambda$ may contain
other highest weights $\Lambda^w$ being, for certain $w$
in the  relevant Weyl group $\W$, shifted (or
affine) Weyl reflections given
by,{\foot{A simple
example is for the ${\V}_{\ell}$ Verma module of $SO(d)$
whereby the state
$|\uell^{\si_{12}}\rangle=(E_{12}^{-+})^{\ell_1-\ell_2+1}|\uell\rangle^{\rhw}$,
 for $\si_{12}(\ell_1,\ell_2,\dots)=(\ell_2,\ell_1,\dots)$,
characterises a sub-Verma-module ${\V}_{\ell^{\si_{12}}}$,
for $|\uell\rangle^{\rhw}$ being annihilated by all $SO(d)$
raising operators.
To see this it suffices to use that $[E_{12}^{+-},(E_{12}^{-+})^{n}]=
4n(H_1-H_2+n-1) (E_{12}^{-+})^{n-1}$.  For unitary
representations $E_{ij}^{\vep\eta}{}^{\dagger}=E_{ij}^{-\vep\,-\eta}$
and $|\uell^{\si_{12}}\rangle$ is null.}}
\eqn\deflambda{
\Lambda^w=w(\Lambda+\rho)-\rho\,,
}
for $\rho$ being the Weyl vector, $\rho=-\half \sum_{\al\in\Phi_-}\al$.
As described more
in Appendix A,
$\V_{\Lambda^w}$ is a sub-Verma module if and only if
the $w$ can be made to satisfy
condition $(A.9)$.
A necessary condition is that
$\Lambda-\Lambda^w$ be expressible as a linear
combination of
positive roots with non-negative integer coefficients. 
This is equivalent  to demanding that the
state with weight $\Lambda^w$  can be
reached by applying
lowering operators on the highest weight state
with weight $\Lambda$.
If the highest
weight $\Lambda$ is dominant integral then all $\V_{\Lambda^w}$ are
sub-Verma modules of $\V_{\Lambda}$.

As described in appendix A, in order to find the character
of $\I_\Lambda$ the first step is to
find all $\Lambda^w$ which are highest weights
of sub-Verma modules.  
Below we give necessary conditions
for $SO^*(d+2)$ weights to satisfy this for
the highest weight having orthonormal basis
labels $\ulambda=(-\De,\ell_1,\dots,\ell_r)$.
We show how each such weight may be
written as $\Lambda'{}^{w}$
where $w\in \W_d$, the Weyl group of $SO(d)$,
and $\Lambda'=(-\De',\uell')$
has $\uell'$ satisfying \piffle\ or \piffles,
so that $\ell'$ is a dominant integral
weight of $SO(d)$, or else such that
the weight $\ell'$ has a Dynkin label
equal to $-1$ in which case
such contributions vanish in $\chi_\Lambda$.
Using the results of appendix A we may then
write the character as
\eqn\projs{
\chi_\Lambda=\sum_{w\in\W_d}{\rm sgn}(w)\C_{\Lambda^w}+
\sum_{w\in\W_d,\,\Lambda'}\gamma_{\Lambda'}{\rm sgn}(w)\C_{\Lambda'{}^{w}}\,,
}
where $\C_{\Lambda'}$
are $SO^*(d+2)$ Verma module
characters and $\gamma_{\Lambda'}$ is determined
by a recurrence relation.  To solve this recurrence
relation requires knowing in more
detail which sub-modules are contained in
which and so condition
(A.9) applies here.

The Weyl group $\W_{d+2}$ acts in a particularly
simple way on weights
of the Verma module ${\V}_\Lambda$ of $SO^*(d+2)$
 in the orthonormal basis.  Choosing any
$w\in \W_{d+2}$ then we may write
\eqn\weylgpacte{
w(\Lambda_0,\dots,\Lambda_r)=
(\vep_0 \Lambda_{\si(0)},\dots,\vep_r \Lambda_{\si(r)})\,,
}
for $\si\in\S_{r+1}$ and $\vep_a=\pm 1$
with $\prod_a \vep_a=1$ for $d=2r$.
In the present case the relevant Weyl vector has components
\eqn\defweyl{
\rho_a=\half d-a\,,\qquad a=0,\dots,r\,,
}
in the orthonormal basis of $SO^*(d+2)$. 
Notice that the last $r$ components are the
components of the Weyl vector for $SO(d)$.
From \deflambda\ we have that
the components of $\Lambda^w$ in the orthonormal
basis are
\eqn\compts{
\Lambda^w_{a}=\vep_a\Lambda_{\si(a)}+(\vep_a-1)\half d-\vep_a \si(a)+a\,.
}

Now for $SO^*(d+2)$ the weights $\ulambda=(-\De,\ell_1,\dots,\ell_r)$
are clearly not dominant integral unless $\De=\ell_i=0$ which
corresponds to the trivial representation.
Sub-Verma-module weights must satisfy \weights.
Thus,
for any $\Lambda^w$ to be
the highest weight of a sub-Verma-module
then  $\Lambda^w_0=-\De-n,\,n\in \Bbb{N}$.
 Also the minimum value of $\Lambda^w_0$ is that for
$\vep_0=-1,\,\si(0)=1$ so that sub-Verma-modules exist
for
\eqn\impo{
-\ell_1-d+1\leq \Lambda^w_0=-\De-n.
}
Notice that for $\vep_0=1,\,\si(0)=0$ so that $\Lambda^w_0=-\De$
then all ${\V}_{\Lambda^w},\,w\in\W_d$ are 
sub-Verma-modules as $\ell$ is a dominant integral 
highest weight with respect to the $SO(d)$ subgroup.
For any other $\vep_0,\si(0)$ then this corresponds
to a definite action of $\P_v$ on the highest
weight state so that for these cases $n\geq 1$.

Using the formula
\projs\ and the results of appendix A with
condition \impo\
we now discuss the even and odd dimension 
character formulae
separately.

\noindent{\bf Character formulae: even dimensions}

For application of the results of appendix A we need to specify
what condition (A.9) demands for $SO^*(d+2)$
Weyl group elements in $d=2r$
dimensions.  For $S_{ab},\,T_{(ab)}$ being the $\W_d$ element
for the respective positive roots 
$\e_a+\e_b,\,\e_a-\e_b,\,0\leq a<b\leq r$ then
clearly
\eqn\clearly{\eqalign{
& S_{ab}(\Lambda_0,\dots,\Lambda_a,\dots,\Lambda_b,\dots,\Lambda_{r})
=(\Lambda_0,\dots,-\Lambda_b,\dots,-\Lambda_a,\dots,\Lambda_r)\,,\cr
& T_{(ab)}(\Lambda_0,\dots,\Lambda_a,\dots,\Lambda_b,\dots,\Lambda_{r})
=(\Lambda_0,\dots,\Lambda_b,\dots,\Lambda_a,\dots,\Lambda_r)\,.
}}
$T_{(ab)}$
corresponds to
the transposition $(ab)$
and below we use the short-hand notation $T_{\si}=T_{(a_1b_1)}\dots
T_{(a_jb_j)}$ for $\si=(a_1b_1)\dots(a_jb_j)$.
Applying $S_{ab}$,
respectively $T_{(ab)}$, to some weight $\ulambda'=(\Lambda'{}_0,\dots
,\Lambda'{}_r)$ then
clearly condition (A.9) 
allows only those $S_{ab}$,
respectively $T_{(ab)}$,
 for which $\Lambda'{}_a+\Lambda'{}_b\in\Bbb{N}$,
respectively $\Lambda'{}_{a}-\Lambda'{}_{b}\in \Bbb{N}$.

We may easily write the character formula for the
$SO^*(2r+2)$
Verma module with highest weight $\Lambda$, for
$\ulambda=(-\De,\uell)$, 
and weights $\Lambda'$
given by \weights\
as
\eqn\charverma{
C^{(2r+2)}_{\ulambda}(s,\x)=
\sum_{\Lambda'}e^{\Lambda'}(\mu)=s^\De\,
C^{(2r)}_{\uell}(\x)\,
P^{(2r)}(s,\x)\,,
}
where, for some general $SO^*(d+2)$ weight
 $\mu$, 
\eqn\defvari{
s=e^{-\e_0}(\mu)=e^{-\mu_0}\,,\qquad x_i=e^{\e_i}(\mu)=
e^{\mu_i}\,,
}
$C^{(2r)}_{\uell}(\x)$ denotes the character of the $SO(2r)$
Verma module with highest weight $\ell$ 
(given in appendix B) and
\eqn\defP{
P^{(2r)}(s,\x)=\prod_{1\leq i\leq r}(1-s x_i)^{-1}(1-s x_i{}^{-1})^{-1}\,.
}
($P^{(2r)}(s,\x)$ 
comes from the summation over $n_{i\pm}$ implicit in \charverma.)

For $\De>\ell_1+d-2$ or
$\De$ lying
between two any of $\De_p$
and $\De_{p+1}$ in \unitarityd\ then \impo\
implies that the only
 sub-Verma-modules are at most
those having highest weights for $\vep_0=1,\,\si(0)=0$
in \compts.
However, since $\ell$ is a dominant integral
weight of $SO(d)$ then all $\V_{\Lambda^w},\,w\in \W_d$
are sub-Verma modules of $\V_{\Lambda}$ in this case.
Thus from \projs\ the corresponding
character is, 
\eqn\defq{
\A^{(2r)}_{[\De;\uell]}(s,\x)
=\sum_{w\in\W_{2r}}{\rm sgn}(w)C^{(2r+2)}_{\ulambda^w}(s,\x)
=s^{\De}\chi^{(2r)}_{\uell}(\x)P^{(2r)}(s,\x)\,.
}

Let us assume that 
$\ell_1=\ell_2=\dots=\ell_p>|\ell_{p+1}|,\,p\leq r-1,\,\De=\De_p$.
In this case only 
$\Lambda^w_0$ for $\vep_0=1,\,\si(0)=0$ and
$\vep_0=-1,\,\si(0)=1,\dots,p$ in \compts\
satisfy \impo.  
For $\vep_0=1,\,\si(0)=0$
then all $\V_{\Lambda^w},\,w\in \W_d$ are sub-Verma
modules.
Let us assume $\vep_0=-1,\,\si(0)=j,\,1 \leq j\leq p$ for
which $\Lambda^w_0=-\ell_1-d+j$ then it
is not difficult to show that the rest of the components
may be written as
\eqn\restcomts{
{(\ell_1,\dots,\ell_1,\ell_{p+1},\dots)^w  =
(\ell_1,\dots,\ell_{1},\ell_1-1,\dots, \ell_1-1,\ell_{p+1},\dots
)^{w'}\,,\quad w'\in\W_d\,,\atop{\,\,\,\,
\uparrow\atop j^{\rm th}\,\,{\rm position}}}
}
whereby if in the original $w\in \W_{d+2}$ then $\si(k_a)=a,\,k_a\neq
0,\,a\neq j$ then
$w'$ is defined in terms of $w$ by
\eqn\defwp{
\vep'_0=-\vep_0=1\,,\quad \vep'_{k_0}=-\vep_{k_0}
\,,\quad \vep'_{k_i}=\vep_{k_i}\,,\quad
\si'=(0\,p\,p-1\dots j)\si\,,
}
so that $\si'(0)=0$ and $\si'$ exhausts 
all members of $\S_r$.
Thus we have shown that all weights for the case
of $w\in \W_{d+2}$ having $\vep_0=-1,\,\si(0)=j$ in \compts\ may be 
written as $\Lambda^{(j,p)}{}^{w'},\,w'\in \W_d$
with{\foot{Also $\prod_a\vep'_a=\prod_a\vep_a$ and
${\rm sgn}(w')=(-1)^{p+j+1}{\rm sgn}(w)$.  During the course
of this work we noticed that,
at this point, simply using the following formula
$${
\chi_\Lambda=\C_{\Lambda}+\sum_{w\in \W,w\neq 1\atop
\Lambda^w\prec\Lambda}{\rm sgn}(w)\C_{\Lambda^w}
}$$
where 
the sum runs over all $w$ satisfying
condition (A.9), gives exactly the same result
for the character as we find here by a more laborious procedure.
It would be interesting to know whether or not this formula
holds for more general Lie algebras and highest weights.
We have not been able to find such a simple
formula in the literature.}} 
\eqn\lambdaj{
\ulambda^{(j,p)}=(-\ell_1-d+j,\ell_1,\dots,\ell_{1},
\ell_1-1,\dots, \ell_1-1,\ell_{p+1},\dots
)\,.\qquad\qquad\atop{
\uparrow\atop (j+1)^{\rm th}\,\,{\rm position}}
}

We may now easily show that $\Lambda^{(j,p)}{}^{w'},\,w'\in \W_d$
exhaust all other highest weights of
sub-Verma-modules of $\V_{\Lambda}$ in this case.
To see this notice that if condition (A.9) of appendix 
A is satisfied for the weight $\Lambda^{(j,p)}$ then
it is satisfied for $\Lambda^{(j,p)}{}^{w'},\,w'\in \W_d$
since the $(\ell_1,\dots,\ell_{1},\ell_1-1,\dots, 
\ell_1-1,\ell_{p+1},\dots
)$ corresponds to a dominant integral weight of $SO(d)$.
For $\Lambda^{(j,p)}$ itself we may show that
$\Lambda^{(j,p)}=\Lambda^w$
where $w=S_{0p}T_{\si}$ for
$\si=(p\,p{-1}\dots j)$ satisfies condition
(A.9).  We see this as  $(p\,p-1\dots j)=(p\,p{-1})(p{-1}\,p{-2})\dots 
(j{+1}\,j)$ 
and $T_{(i{+1}\,i)}T_{(i\,i{-1})}\dots
T_{(j{+1}j)}(\ulambda+{\underline \rho})
\cdot (\ue_{i{+2}}-\ue_{i{+1}})\in \Bbb{N} $ for
$i<p$ and  
$T_{\si}(\ulambda+{\underline \rho})\cdot(\ue_0+\ue_p)\in \Bbb{N} $.

It now remains to determine $\gamma_{\Lambda^{(j,p)}}\to\gamma_{j,p}$ 
in \projs\ for the
weight $\Lambda^{(j,p)}$ by the recurrence 
relation in appendix A.
The first observation which is not difficult to
show (in a similar fashion
as above) is that among $\V_{\Lambda^{(k,p)}}$ the
only such modules which contain $\V_{\Lambda^{(j,p)}}$
as a sub-module are those for $j<k\leq p$
or none such if $j=p$.
In fact it is possible to show that 
no proper sub-module of $\V_{\Lambda}$
contains $\V_{\Lambda^{(p,p)}}$
so that in this case $\gamma_{p,p}=-1$.

To describe which sub-modules
contain other $\V_{\Lambda^{(j,p)}}$ we first
define a subset of the permutation group
${\T}_n\subset{\S}_n$  
so that every $\tau\in{\T}_n$ has $1\leq c\leq n$ cycles,
where the first cycle consists of the first $n_1$
of the integers $n,n{-1},\dots,2,1$, preserving this ordering,
the second consists of the next $n_2$ such integers,
again preserving this ordering, and so on and where 
$n_1,\dots,n_c\geq 1$ satisfy $\sum_{i=1}^c n_i=n$.
For example for $n=3$ then 
${\T}_n=\{(321),(32)(1)=(32),(3)(21)=(21),(3)(2)(1)=1\}$.
It is not difficult to see that the number of
such permutations with $c$ cycles (counting trivial
one cycles) is $\Big({n-1\atop c-1}\Big)$
so that the total number of such permutations is
$2^{n-1}$.  Further we have that for $n>1$ there
are $2^{n-2}$ of these permutations with
signature $1$ or $-1$ so that $\sum_{\tau\in{\T}_n}{\rm sgn}(\tau)=0$.

With the above definition of ${\T}_n$ we have found that
the sub-modules
$\V_{\Lambda^{w'}},\,w'\in \W_d,w'\neq 1$ contain $\V_{\Lambda^{(j,p)}}$
as a sub-module only for $w'\in {\T}_{p+1-j}$
so that for $j>p$ then $\sum_{w'}{\rm sgn}(w')=-1$
is the contribution to $\gamma_{j,p}$ coming from these.
Also we have found that the sub-modules $\V_{\Lambda^{(k,p)}{}^{w'}},\,
j<k\leq p,\,w'\in \W_d$ containing $\V_{\Lambda^{(j,p)}}$
have $w'\in {\T}_{k-j}$ so that for $j+1<k\leq p$
then $\sum_{w'}{\rm sgn}(w')=0$ so that the contribution
to  $\gamma_{j,p}$ coming from these vanishes
while for $k=j+1$ then $w'=1$ and this contributes
$\gamma_{j+1,p}\,{\rm sgn}(w')=\gamma_{j+1,p}$ to
$\gamma_{j,p}$.
With these results we may then easily find that
\eqn\gammadef{
\gamma_{j,p}=(-1)^{p+j+1}\,,
}
solves the recurrence relation 
$\gamma_{j,p}=-\gamma_{j+1,p},\,\gamma_{p,p}=-1$.

It is possible to show that for such $w'$ as described 
above (essentially belonging to ${\T}_n$ for various $n$)
then for
$w\in \W_{d+2}$ so that $\Lambda^{(j,p)}=\Lambda^{ww'}$ or
$\Lambda^{(j,p)}=\Lambda^{(k,p)}{}^{ww'}$
the only $w$ which satisfy
condition (A.9) are expressible
as products of $S_{0l},\,T_{(mn)}$, $1\leq l,m,n\leq p$.

Now applying \projs\ and the Weyl character formula for
$SO(2r)$ in terms of \charverma\ (appendix B)
we find that the corresponding
character is,
\eqn\defr{\eqalign{
&{} \D^{(2r)}_{[\ell_1+2r-p-1;\ell_1,\ell_{p+1},\dots,\ell_r]}(s,\x)\cr
&{}=s^{\ell_1+2r-p-1}\Big(\chi^{(2r)}_{\uell}(\x)
+\sum_{1\leq j\leq p}(-s)^{p+1-j}
\chi^{(2r)}_{\uell-\ue_p-\dots-\ue_j}(\x)\Big)P^{(2r)}(s,\x)\,.}
}
 for $\uell=(\ell_1,\dots,\ell_1,\ell_{p+1},\dots,\ell_r)$.

Notice that for even dimensions we also have the possibility
of $\ell_1=\dots=\ell_{r-1}=\pm\ell_r,\,\De=\ell_1+\half d-1$.
Here the $\Lambda^w_0$ satisfying \impo\ are those for $\vep_0=1,\,\si(0)=0$,
$\vep_0=\mp 1,\,\si(0)=r$ along with $\vep_0=-1,\,\si(0)=1,\dots,r-1$.
By an argument very similar to the previous we find that,
for $\uell=(\ell,\dots,\ell,\pm \ell)$,
\eqn\defs{\eqalign{
&{}\D^{(2r)}_{[\ell+r-1;\ell]\pm}(s,\x)\cr
&{}=s^{\ell+r-1}\Big(\chi^{(2r)}_{\uell}(\x)+\sum_{1\leq j\leq r}(-s)^{r+1-j}
\chi^{(2r)}_{\uell\mp\ue_r-\dots-\ue_j}(\x)\Big)P^{(2r)}(s,\x)\,,}
}
is the corresponding character in this case.

The free scalar case, for which $\ulambda=(-r+1,0,\dots,0)$ is
the highest $SO(2r+2)$ weight,
is accounted for by \defs\ for $\ell=0$ and has character
\eqn\defsr{
\D^{(2r)}_{[r-1;0]}(s,\x)\equiv\D^{(2r)}_{[r-1;0]\pm}(s,\x)
=s^{r-1}(1-s^2)P^{(2r)}(s,\x)\,,
}
since the $SO(2r)$ characters obey
$\chi_{(0,\dots,0,-1,\pm 1)}(\x)=-1$ with 
$\chi_{(0,\dots,0,\pm 1)}(\x)=0$ and
$\chi_{(0,\dots,0,-1,\dots,-1,\pm
1)}(\x)=0$
otherwise.

\noindent
{\bf Character formulae: odd dimensions}

Considerations for odd $d=2r+1$ dimensions
are very similar to those above for even $d=2r$ dimensions
and we will not go into as much detail here. 
Along with \clearly\ $\W_{2r+1}$ has extra elements
corresponding to the extra positive roots $\e_{i},\,1\leq i\leq r$ 
given by
\eqn\extar{
S_{a}(\Lambda_0,\dots,\Lambda_a,\dots\Lambda_r)=
(\Lambda_0,\dots, -\Lambda_a,\dots,\Lambda_r)\,.
}
Acting on some weight $\ulambda'=(\Lambda'{}_0,\dots,\Lambda'{}_r)$
then condition (A.9) only allows those $S_a$ for which $\Lambda'{}_a\in
{1\over 2}\Bbb{N}$.

We may easily write the character formula for the
$SO^*(2r+3)$
Verma module with highest weight $\Lambda$ and weights $\Lambda'$
given by \weights\
as
\eqn\charvermao{
C^{(2r+3)}_{\ulambda}(s,\x)=
\sum_{\Lambda'}e^{\Lambda'}(\mu)=s^\De\,
C^{(2r+1)}_{\uell}(\x)\,
P^{(2r+1)}(s,\x)\,,
}
where
$C^{(2r+1)}_{\uell}(\x)$ denotes the character of the $SO(2r+1)$
Verma module with highest weight $\ell$ 
(given in appendix B) and
\eqn\defPo{
P^{(2r+1)}(s,\x)=(1-s)^{-1}
\prod_{1\leq i\leq r}(1-s x_i)^{-1}(1-s x_i{}^{-1})^{-1}\,.
}
($P^{(2r+1)}(s,\x)$ 
comes from the summation over $n_{i\pm},n_0$ implicit in \charvermao.)

For $\De>\ell_1+d-2$ or
$\De$ lying
between two any of $\De_p$
and $\De_{p+1}$ in \unitarityd\ then
character for positive energy unitary irreducible
representations is, 
\eqn\defqo{
\A^{(2r+1)}_{[\De;\uell]}(s,\x)
=\sum_{w\in\W_{2r+1}}{\rm sgn}(w)C^{(2r+3)}_{\ulambda^w}(s,\x)
=s^{\De}\chi^{(2r+1)}_{\uell}(\x)P^{(2r+1)}(s,\x)\,,
}
 where $\chi^{(2r+1)}_{\uell}(\x)$ is
the character of the $SO(2r+1)$ irreducible representation with
highest weight $\ell$.

For $\De=\De_p$ in
\unitarityd\
we may go through the same procedure
as for the even dimensional case
and find that the extra Weyl reflections \extar\
lead to nothing new
as far as condition (A.9)
is concerned. Thus for odd $d$ and with 
$\uell=(\ell_1,\dots,\ell_1,\ell_{p+1},\dots,\ell_r)$
the corresponding
character for these representations is,
\eqn\defro{\eqalign{
&{} \D^{(2r+1)}_{[\ell_1+2r-p;\ell_1,\ell_{p+1},\dots,\ell_r]}(s,\x)\cr
&{}=s^{\ell_1+2r-p}\Big(\chi^{(2r+1)}_{\uell}(\x)
+\sum_{1\leq j\leq p}(-s)^{p+1-j}
\chi^{(2r+1)}_{\uell-\ue_p-\dots-\ue_j}(\x)\Big)P^{(2r+1)}(s,\x)\,.}
}

For the free scalar case,
for which $\ulambda=(-r+{1\over 2},0,\dots,0)$
is the highest $SO(2r+3)$ weight, we have that the corresponding character is
given by,
\eqn\deft{
\D^{(2r+1)}_{[r-{1\over 2};0]}(s,\x)
=s^{r-{1\over 2}}(1-s^2)P^{(2r+1)}(s,\x)\,.}

For odd dimensions we also
have the possibility of the highest
weight $\Lambda$ having components $\ulambda=(-r,{1\over 2},\dots
{1\over 2})$.  This time the $\Lambda^w_0$ satisfying
\impo\ are those for $\vep_0=\pm 1,\,\si(0)=0$,
$\vep_0=-1,\,\si(0)=j$,
$1\leq j \leq r$.  For $\vep_0=-1,\,\si(0)=0$ then $\Lambda^w_0=-r-1$
and the remaining components may be rewritten as $\uell^w=\uell^{w'}$
where $w'\in\W_{2r+1}$ is identical to $w$ save for $\vep_0'=-\vep_0=1$ so that
$\prod_a\vep_a'=-\prod_a\vep_a$ and ${\rm sgn}(w')=-{\rm sgn}(w)$.
The cases of
$\vep_0=-1$,
$\si(0)=j$, $1\leq j\leq r$ are accounted for similarly
as for even dimensions for $p=r$.
However these cases have highest weights which are
shifted $\W_{2r+1}$ Weyl group reflections of
$(-\De',\uell')=(-2r-1+j,-{1\over 2},\dots, -{1\over 2},{1\over 2},\dots)$.
For these weights at least one of the Dynkin labels 
$\ell'{}_{i+1}-\ell'{}_{i}$
is equal to $-1$ so that contributions 
from all these Verma modules vanish from the
character formula (by a result of appendix A).

The only cases we have to consider are for the Verma modules
$\V_{\Lambda'{}^{w'}},\,w'\in \W_{2r+1}$ for
$\ulambda'=(-r-1,{1\over 2},\dots,{1\over 2})$.
By similar arguments as before all $\V_{\Lambda'{}^{w'}}$
for $w'\neq 1$ are sub-modules of
$\V_{\Lambda'}$.  Due to $\ulambda'+{\underline\rho}
=S_0(\ulambda+{\underline \rho})$, where $S_a$ is defined in \extar, then
 condition (A.9) (which is satisfied due to
$\e_0\cdot (\ulambda+{\underline \rho})={1\over 2}$)
implies that all $\V_{\Lambda'{}^{w'}},\,w'\in \W_{2r+1}$
are sub-modules of $\V_{\Lambda}$.  Further we may show that
$\V_{\Lambda'}$ is contained only in $\V_{\Lambda}$,
and no sub-modules of $\V_{\Lambda}$, and  so 
$\gamma_{\Lambda'}=-1$.  

Taking into account these considerations, 
we have that the character is, from \projs,
\eqn\defto{\eqalign{
\D^{(2r+1)}_{[r;{1\over 2}]}(s,\x)&{}=\sum_{w\in \W_{2r+1}}{\rm
sgn}(w)\big(
\C_{\Lambda^{w}}-\C_{\Lambda'{}^{w}}\big)\cr
&{}
=s^{r}(x_1{}^{1\over 2}+x_1{}^{-{1\over 2}})
\dots (x_r{}^{1\over 2}+x_{r}{}^{-{1\over
2}})(1-s)P^{(2r+1)}(s,\x)\,,}
}
and where we have used that
\eqn\deftexo{
\chi^{(2r+1)}_{({1\over 2},\dots,{1\over 2})}(\x)
=(x_1{}^{1\over 2}+x_1{}^{-{1\over 2}})
\dots (x_r{}^{1\over 2}+x_{r}{}^{-{1\over 2}})\,.
}

\noindent
{\bf{Relation with reduced Verma module bases}}

We wish to show here that omitting certain of $\P_v$
from the original Verma module basis
\cotwentythree\ leads to formulae for
characters which are equivalent to those
obtained in the last sections.

Descendant states 
(i.e. those obtained 
by definite action of $\P_v$
on the highest weight state) are
$SO(d)$ representations belonging to 
the decomposition of $\e_1\otimes \dots\otimes \e_1\otimes \ell$
in terms of irreducible representations.
From appendix C, these states are null for either of two reasons.  
One reason is that $\ell$ may lie on the boundary
of the dominant Weyl chamber \piffle\ (i.e. that
some $\ell_1=\ell_2=\dots$) so that certain descendant states
are null with respect to the $SO(d)$ subgroup.
The other reason is that $\De$ may lie on a unitarity
bound.  By omitting the correct $\P_v$ from \cotwentythree\
we effectively discard states in the original Verma
module which are null due to the value of $\De$.
Acting with the Weyl symmetry operator on
the character of the reduced Verma module is equivalent
to projecting 
out of the reduced module states which are null by
virtue of which $SO(d)$ representation they belong to.
We may show that this prescription gives the same
formulae for characters as found earlier.
% The result is the character of a uni

For definiteness we consider the case where 
$\ell_1=\dots=\ell_p>|\ell_{p+1}|,\,\De\geq\ell_1+d-p-1$,
in even dimensions, $d=2r$, although the other cases
of interest are similar.

For this case and
in the notation of appendix
C, consider the $SO(2r)$ highest weight state $|\De+1,\uell-\ue_p\r$.
The construction of such a state is non-trivial and
in appendix C only the simplest such states have been
constructed.  
Nevertheless we may write down the state in principle
as
\eqn\stste{
|\De+1,\uell-\ue_p\r=\A_{\uell}\,\P_{p-}|\De,\uell\r^{\rhw}+
\sum_{v',\uell'}
\B_{v',\uell'}\P_{v'}|\De,\uell'\r\,,
}
where $\P_{v'}\leftrightarrow \v'\in\{\pm \e_i\}$ is the 
subset of
$\P_{i\pm}$ which may be reached by applying $SO(2r)$ 
raising operators to $\P_{p-}$ and
$\uell'+\uv'=\uell-\ue_p$.  The subset $\P_{v'}$ may be
easily determined from \weightsystemeven\
to be given by $\P_{i+},\P_{j-}$ 
for $1\leq i\leq r$ and $p+1\leq j\leq r$.
The complex numbers $\A_{\uell},\,\B_{v',\uell'}$
are determined by the condition that \stste\ be
a highest weight state with respect to $SO(2r)$
i.e. that all $SO(2r)$ raising operators annihilate it.
By the results of appendix C,
for $\De$ above the unitarity bound this state is not null
however when $\De=\ell_1+d-p-1$ 
then $|\De+1,\uell-\ue_p\r=0$. This is
equivalent to a conservation equation for the
highest weight state $|\De,\uell\r^{\rhw}$.

Now the modulus of $\A_{\uell}$ is non-zero
despite $\ell_1=\dots=\ell_p>|\ell_{p+1}|$.
When $\De=\ell_1+d-p-1$, so that \stste\ vanishes,
 then $\P_{p-}|\De,\uell\r^{\rhw}$
may be expressed in terms of $\P_{i+}|\De,\ell'\r,\,
\P_{j-}|\De,\ell'\r$
for $1\leq i\leq r$ and  $p+1\leq j\leq r$.
Also, as 
\eqn\importante{
E^{-+}_{i\,i+1}
|\De,\ell_1,\dots,\ell_1,\ell_{p+1},\dots,\ell_r\r^{\rhw}=0\,,\quad
i=1,\dots,p-1\,,
}
then applying such $E^{-+}_{i\,i+1}$ to
\stste\ and using \weightsystemeven,  
we have that for $\De$ on the unitarity bound 
then $\P_{i-}|\De,\uell\r^{\rhw},\,
1\leq i\leq p-1$ may be similarly 
expressed in terms of $\P_{i+}|\De,\ell'\r,\,
\P_{j-}|\De,\ell'\r$
for $1\leq i\leq r$ and  $p+1\leq j\leq r$.
Thus, effectively the Verma module basis \cotwentythree\
becomes reduced so as to exclude $\P_{i-},\,1\leq i\leq p$.

Acting with the $SO(d)$ Weyl symmetry 
operator $\frak{W}_d$ (see appendices A,B) on the
character for the reduced Verma module
yields the following formula,
\eqn\charspeccase{\eqalign{
\sum_{\Lambda',
w\in \W_{2r}}e^{w(\Lambda')}(\mu)&{}
=s^{\ell_1+2r-p-1} {\frak{W}}_d\Big(C^{(2r)}_{\uell}(\x)
(1-s x_{1}{}^{-1})\dots (1-s x_{p}{}^{-1})\Big)\,P^{(2r)}(s,\x)\cr
&{} =s^{\ell_1+2r-p-1}\Big(\chi^{(2r)}_{\uell}(\x)+\!\!\!\!\!\!\!\!
\sum_{1\leq n\leq p\atop i_j\in\{1,\dots,p\},\,i_j\neq i_k}\!\!\!\!\!\!\!
(-s)^{n}\chi^{(2r)}_{\uell-\ue_{i_1}-\dots-\ue_{i_n}}(\x)\Big)
\,P^{(2r)}(s,\x)\,,}
}
for $\uell=(\ell_1,\dots,\ell_1,\ell_{p+1},\dots,\ell_r)$
and
where now $\ulambda'=(-\Lambda'_0,\uell')$ are specified  by
\eqn\newlambdap{\eqalign{
\Lambda'_0 &{} =-\De'=-\ell_1-d+p+1-\sum_{1\leq
i\leq r}n_{i+}-\sum_{p+1\leq j\leq r}n_{j-}\,, \cr
\ell'&{}=\ell-\sum_{\al\in\Phi_-}\al+\sum_{1\leq
i\leq r}n_{i+}\e_i-\sum_{p+1\leq j\leq r}n_{j-}\e_j\,.
}}
It is easy to see that \charspeccase\ reduces to
\defr. To see this note that
$\chi_{\uell'}(\x)$ in \charspeccase\ is non-zero
only for
$\uell'=(\ell_1,\dots,\ell_1,\ell_1-1,\dots,
\ell_1-1,\ell_{p+1},\dots,\ell_r)$ i.e. for $i_j=k,k+1,\dots,p$
for some $1\leq k\leq p$.

%A plausible, but not rigorously proven, reason
%for why this procedure yields the same formula as before
%is as follows.

%of the conformal group which is 
%irreducible.

To summarise: when the conformal dimension $\De$ saturates a unitarity
bound the Verma module basis is reduced so as to
exclude certain of the $\P_{i\pm},\,\P_0$ from
\cotwentythree.  This is equivalent to conservation
equations constraining the highest weight state.
This subset may be determined
in terms of the results of appendix C.  
  Acting
with the $SO(d)$ Weyl symmetry operator on
the character of the reduced Verma module then leads to 
the characters for the corresponding unitary irreducible
representation. Explicitly,
for \defr\ or \defro\ the subset to be omitted
from \cotwentythree\ is $\P_{i-},\,1\leq i\leq p$,
for \defs\ the subset is $\P_{i-},\,1\leq i\leq r-1$
along with $\P_{r-}$ for $\D^{(2r)}_{[\ell+r-1;\ell]+}$
or $\P_{r+}$ for $\D^{(2r)}_{[\ell+r-1;\ell]-}$ while
for \defto\ the subset is $\P_{i-},\,1\leq i\leq r$ along
with $\P_0$.

\noindent
{\bf Special cases}

We here illustrate these character formulae for the simplest cases of
the $SO(3,2)$ and $SO(4,2)$ conformal groups and
mention how special cases relate to conformal field representations.  

We have that,
\eqn\cotwentyten{
C_\ell(x)\equiv C^{(3)}_{\ell}(x)={x^\ell\over 1-x^{-1}}\,,
}
is the character of $SO(3)$ Verma modules and
\eqn\cotwentynine{
\chi_{\ell}(x)\equiv \chi^{(3)}_{\ell}(x)
=C_\ell(x)+C_\ell(x^{-1})={x^{\ell+{1\over
2}}-x^{-\ell-{1\over 2}}\over x^{1\over 2}-x^{-{1\over 2}}}\,,
}
is the usual character for $SO(3)$ irreducible representations.
We have therefore that,{\foot{
In terms of the notation employed in \Flato\
these character
formulae agree for
$x\to\be^2,\,s\to\al^2$. 
In the nomenclature of \Flato, 
the representations $\D^{(3)}_{[1;{1\over 2}]}$ and 
$\D^{(3)}_{[{1\over
2};0]}$
correspond to the `Di' and `Rac'
singleton representations, respectively.}}
\eqn\cotwentyeight{\eqalign{
\A^{(3)}_{[\De;\ell]}(s,x)&{}
=s^{\De}\,\chi_{\ell}(x)P^{(3)}(s,x)\,,\cr
\D^{(3)}_{[\ell+1;\ell]}(s,x)&{}
=s^{\ell+1}\,\left(\chi_{\ell}(x)-
s \chi_{\ell-1}(x)\right)P^{(3)}(s,x)\,,\cr
\D^{(3)}_{[1;{1\over 2}]}(s,x)&{}
=s\,(x^{1\over 2}+x^{-{1\over 2}})\,(1-s x)^{-1}(1-s x^{-1})^{-1}\,,\cr
\D^{(3)}_{[{1\over 2};0]}(s,x)
&{}=s^{{1\over 2}}\,(1+s)\,(1-s x)^{-1}(1-s x^{-1})^{-1}\,,
}}
exhaust all characters of the unitary irreducible
representations of $SO(3,2)$.  Analogous formulae are to be
found in \dobo.

For $SO(4)\simeq SU(2)\otimes SU(2)$ we have that 
\eqn\charsoft{
C^{(4)}_{(\ell_1,\ell_2)}(x_1,x_2)=C_j(x)C_{\bj}(y)\,,\,\,\, {\rm for}
\,\,\,  \ell_1=j+\bj\,,
\quad \ell_2=j-\bj\,,\quad x_1=x^{{1\over 2}}y^{{1\over 2}}\,,
\quad x_2=x^{{1\over 2}}y^{-{1\over 2}}\,,
}
i.e. the Verma module character with dominant highest weight
$(\ell_1,\ell_2)$ may be expressed as a product of two $SU(2)$
Verma module characters with highest weights $j,\,\bj$.
The characters of
unitary irreducible
representations of $SO(4,2)$
are given by,
\eqn\cten{\eqalign{
\A^{(4)}_{[\De;j,\bj]}(s,x,y)&{} =s^\De
\chi_j(x)\chi_{\bj}(y)P^{(4)}(s,x,y)
\,,\cr
\D^{(4)}_{[j+\bj+2;j,\bj]}(s,x,y)
&{} =s^{j+\bj+2}\left(\chi_j(x)\chi_{\bj}(y)-s\,\chi_{j-{1\over
    2}}(x)\chi_{\bj-{1\over 2}}(y)\right)P^{(4)}(s,x,y)\,,\cr
\D^{(4)}_{[j+1;j]+}(s,x,y)& =s^{j+1}\left(\chi_j(x)-s\,
\chi_{j-{1\over
    2}}(x)\chi_{1\over 2}(y)+s^2\chi_{j-1}(x)\right)P^{(4)}(s,x,y)\,,\cr
{\D}^{(4)}_{[\bj+1;\bj]-}(s,x,y)&  =s^{\bj+1}
\left(\chi_{\bj}(y)-s\,\chi_{\bj-{1\over
    2}}(y)\chi_{1\over 2}(x)+s^2\chi_{\bj-1}(y)\right)P^{(4)}(s,x,y)\,.\cr}
}
Here we have written $P^{(4)}(s,x_1,x_2)\to P^{(4)}(s,x,y)$
for $x_1,x_2$ as in \charsoft.
This reproduces the results for character
formulae in four dimensions found in \Bara.
Similar formulae are to be found in \dobtw.

Free fields have conformal dimension $\ell+{1\over 2}d-1$ and belong
to the $(\ell,\dots,\ell,\pm \ell)$ representation of $SO(2r)$
for any $\ell\in {1\over 2}\Bbb{N}$ in
$d=2r$ dimensions and the $(\ell,\dots,\ell,\ell)$ 
representation
of $SO(2r+1)$ for  $\ell=0,{1\over 2}$ in $d=2r+1$
dimensions \sie.  
The corresponding characters for even dimensions are
$\D^{(2r)}_{[\ell+r-1;\ell]\pm}(s,\x)$ in \defs\
along with $\D^{(2r)}_{[r-1;0]}(s,\x)$ in \defsr\ for the
scalar case.
For odd dimensions the corresponding characters are
$\D^{(2r+1)}_{[r;{1\over 2}]}(s,\x)$ of
\defto\ and $\D^{(2r+1)}_{[r-{1\over 2};0]}(s,\x)$ in \deft.

The characters \defr, \defro\ for the
special case of 
$p=1$ and  $\ell_1\equiv \ell,\,\ell_2=\dots=\ell_r=0$
correspond to conserved symmetric traceless 
tensor-field representations of
the conformal group, $T_{\mu_1\dots\mu_\ell}=
T_{(\mu_1\dots \mu_\ell)},\,T^{\mu}{}_{\mu\mu_2\dots\mu_\ell}=
\pr^{\mu_1} T_{\mu_1\dots\mu_\ell}=0$.
These have conformal dimension $d+\ell-2$ in $d$ dimensions
and examples are the conserved vector current for $\ell=1$
and energy momentum tensor for $\ell=2$.

\newsec{Product formulae}
We now turn to the determination
of the decomposition
of products of unitary irreducible representations of the 
conformal group into other unitary irreducible representations.

\noindent{{\bf Product formulae: four dimensions}}

We illustrate for the $SO(4,2)$ case first.
For these purposes we first note a useful identity, namely,
\eqn\ctwelve{
P^{(4)}(s,x,y)=\sum_{p,q=0}^{\infty}s^{2p+q} \chi_{{1\over 2}q}(x)
\chi_{{1\over 2}q}(y)\,.
}
With \ctwelve\ we may now easily determine the products of unitary
irreducible representations of the conformal group.  Using the
usual decomposition of products of $SU(2)$ characters,
\eqn\cfifteen{
\chi_{j\otimes j'}(x)\equiv \chi_{j}(x)\chi_{j'}(x)=
\sum_{q=|j-j'|}^{j+j'}\chi_q(x)\,,}
and \ctwelve, we notice that
\eqn\csixteen{
\D^{(4)}_{[j+1;j]+}(s,x,y)=\sum_{q=0}^{\infty}s^{q+j+1}\chi_{j+{1\over
    2}q}(x)
\chi_{{1\over 2}q}(y).
}
Thus we may straightforwardly determine that,
\eqn\cseventeen{\eqalign{
&{}\D^{(4)}_{[j+1,j]+}(s,x,y)\D^{(4)}_{[j'+1;j']+}(s,x,y) \cr
&{}=
s^{j+j'+2}P(s,x,y)\bigg(\chi_{j\otimes
  j'}(x)+\sum_{q=1}^{\infty}s^q\big(\chi_{j+j'+{1\over
    2}q}(x)\chi_{{1\over 2}q}(y)-s\,\chi_{j+j'+{1\over
    2}q-{1\over 2}}(x)\chi_{{1\over 2}q-{1\over 2}}(y)\big)\bigg)\cr
&=\A^{(4)}_{[j+j'+2;j\otimes j',0]}(s,x,y)+\sum_{q=1}^{\infty}
\D^{(4)}_{[j+j'+q+2;j+j'+{1\over 2}q,{1\over 2}q]}(s,x,y)\,.}
}
Similarly, using \cfifteen, \csixteen, we may easily determine that,
\eqn\ceighteen{\eqalign{
\D^{(4)}_{[j+1,j]+}&(s,x,y)
\D^{(4)}_{[\bj+1;\bj]-}(s,x,y) \cr
&=
s^{j+\bj+2}P(s,x,y)\sum_{q=0}^{\infty}s^q\big(\chi_{j+{1\over
    2}q}(x)\chi_{\bj+{1\over 2}q}(y)-s\,\chi_{j+{1\over
    2}q-{1\over 2}}(x)\chi_{\bj+{1\over 2}q-{1\over 2}}(y)\big)\cr
&=\sum_{q=0}^{\infty}
\D^{(4)}_{[j+\bj+q+2;j+{1\over 2}q,\bj+{1\over 2}q]}(s,x,y)\,.}
}
Using \cfifteen, \csixteen, we may also find that
\eqn\cnineteen{\eqalign{
& \D^{(4)}_{[j+\bj+2;j,\bj]}(s,x,y)\D^{(4)}_{[j'+1,j']+}(s,x,y)\cr
& =\sum_{q=0}^{\infty}\big(\A^{(4)}_{[\De_q,j+j'+{1\over 2}q,\bj+{1\over
	  2}q]}(s,x,y)+\A^{(4)}_{[\De_q,(j-{1\over 2})\otimes (j'+{1\over
      2}q-{1\over 2}),\bj+{1\over 2}q]}(s,x,y)\cr
&\qquad +\A^{(4)}_{[\De_q,j+j'+{1\over
      2}q,(\bj-{1\over 2})\otimes({1\over 2}q-{1\over
	  2})]}(s,x,y)\big)
\,,\qquad\De_q=j+j'+\bj+q+3\,,}
}
and
\eqn\cnineteen{
 \A^{(4)}_{[\De;j,\bj]}(s,x,y)\D^{(4)}_{[j'+1,j']+}(s,x,y)
= \sum_{q=0}^{\infty}
\A^{(4)}_{[\De+j'+q+1;j\otimes(j'+{1\over 2}q),\bj\otimes{1\over
2}q]}
(s,x,y)\,,
}
which exhausts all products involving $\D^{(4)}_{[j+1;j]+}(s,x,y)$.
Those involving ${\D}^{(4)}_{[\bj+1;\bj]-}(s,x,y)$ may be
obtained by the exchange $x\leftrightarrow y$
above noting
\eqn\canother{\eqalign{
{\D}^{(4)}_{[\bj+1;\bj]-}(s,x,y)&{}=
\D^{(4)}_{[\bj+1;\bj]+}(s,y,x)\,,\quad
\D^{(4)}_{[j+\bj+2;j,\bj]}(s,x,y)=
\D^{(4)}_{[j+\bj+2;\bj,j]}(s,y,x)
\,,\cr
&\qquad\qquad \A^{(4)}_{[\De;j,\bj]}(s,x,y)=\A^{(4)}_{[\De;\bj,j]}(s,y,x)
\,.}
}

Similarly, using \ctwelve\ directly, we have that
\eqn\ctwenty{
\A^{(4)}_{[\De;j,\bj]}(s,y,x)\A^{(4)}_{[\De';j',\bj']}(s,y,x)
=\sum_{p,q=0}^{\infty}
\A^{(4)}_{[\De+\De'+2p+q;j\otimes j'\otimes {1\over 2}q,
\bj\otimes\bj'\otimes {1\over
  2}q]}(s,x,y)\,.
}
We may note that each of these product formulae is compatible with
the `blind' partition functions,
\eqn\ctwenty{\eqalign{
\A^{(4)}_{[\De;j,\bj]}(s,1,1)&={s^{\De}\over (s-1)^4}(2j+1)(2\bj+1)\,,\cr
\D^{(4)}_{[j+\bj+2;j,\bj]}(s,1,1)&={s^{j+\bj+2}\over
  (s-1)^4}\big((2j+1)(2\bj+1)-4 s j \bj\big)\,,\cr
\D^{(4)}_{[j+1;j]+}(s,1,1)& ={\D}^{(4)}_{[j+1;j]-}(s,1,1)
={s^{j+1}\over
  (s-1)^3}\big(-(2j+1)+s(2j-1)\big)\,.}
}

As we see, the $d=4$ cases are relatively simple
when we use expansion formulae of the type \ctwelve,
\csixteen\ to expand one of the characters
in the product of two.  The general cases which we consider
now are also made simpler with analogous expansion
formulae.

\noindent{\bf Product formulae: even dimensions }

Useful for finding product formulae for
the $SO(2r,2)$ conformal group is the following
expansion of $P(s,\x)$ in terms of $SO(2r)$ characters, namely,
\eqn\findthatb{
P^{(2r)}(s,\x)
=\sum_{p,q=0}^{\infty}s^{2p+q}\chi^{(2r)}_{(q,0,\dots,0)}(\x)\,,
}
where
\eqn\findthatc{
\chi^{(2r)}_{(q,0,\dots,0)}(\x)=\half {\rm det}[x_i{}^{k_j}+x_i{}^{-k_j}]
\De(x_1+x_1{}^{-1},\dots,x_r+x_r{}^{-1})^{-1}\,,
}
with $k_1=q+r-1,\,k_j=r-j,\,j>1$, 
for $\De(\x)$ being the Vandermonde determinant,
\eqn\cthirteeno{
\De(x_1,\dots,x_n)=\prod_{1\leq i<j\leq n}(x_j-x_i)\,.
}
The latter expression for the character comes from appendix B
where also expressions for more general characters of $SO(d)$
in even and odd dimensions are given.

Analogously to \csixteen\ we have for \defs\ that
\eqn\csixteenanal{
\D^{(2r)}_{[\ell+r-1;\ell]\pm}(s,\x)
=\sum_{q=0}^{\infty}
s^{\ell+r+q-1}\chi^{(2r)}_{(\ell+q,\ell,\dots,\ell,\pm\ell)}(\x)\,,
}
which we prove in appendix D.

More generally for the
$p=r-j$ case of \defr\ (note that \csixteenanal\ encapsulates the
$p=r$ case) we have, for $\ell>\ell_1>\dots >|\ell_j|$,
\eqn\csixteenanalopo{\eqalign{
& \D^{(2r)}_{[\ell+r+j-1;\ell,\ell_1,\dots,\ell_j]}(s,\x)\cr
&{} =\!\!\!\sum_{p_1,\dots,p_j,q=0}^{\infty}
\sum_{i_1=-{1\over 2}p_1}^{{1\over 2} p_1}\!\dots
\!\sum_{i_j=-{1\over 2}p_j}^{{1\over 2} p_j}\!\!\!
s^{\ell+r+j+q+p_1+\dots+ p_j-1}
\chi^{(2r)}_{(\ell+q,\ell,\dots,\ell,\ell_1+2i_1,\dots,\ell_j+2 i_j)}(\x)\,,}
}
which we also prove in appendix D.
Note that the weights
$(\ell+q,\ell,\dots,\ell,\ell_1+2i_1,\dots,\ell_j+2 i_j)$
may lie outside the dominant Weyl chamber, i.e. not satisfy
\piffle,  for particular 
$\ell,\ell_i$.  However for such weights we may use that
$\chi^{(2r)}_{\uell'}(\x)={\rm sgn}(w)\chi^{(2r)}_{\uell'^{w}}(\x)$,
for some $w\in \W_{2r}$,
to relate such characters to the character with
dominant integral highest weight $\uell'^{w}$.

A notable simplification to \csixteenanalopo\ occurs for the
$p=1$ case of \defr\ for $\ell_1\equiv \ell,\,\ell_2=\dots=\ell_r=0$
which corresponds to conserved symmetric traceless 
tensor representations of
the conformal group.  In this case we obtain that, for $d>4$,
\eqn\anothersimp{
\D^{(2r)}_{[\ell+2r-2;\ell,0,\dots,0]}(s,\x)=\sum_{p,q=0}^{\infty}
\sum_{k=0}^{\ell}s^{\ell+2r+2p+q-2}\chi^{(2r)}_{(q+k,\ell-k,0,\dots,0)}(\x)\,.
}

We now discuss products involving the 
representations in \csixteenanal\ which contain the 
`truncated' representations in \defr\ for $p=1$
namely,
$\D^{(2r)}_{[\ell_1+2r-2,\ell_1,\dots,\ell_r]}$.

For $d=6$, for example, we may find using \csixteenanal, \defr\ for $p=1$ 
and \defs\ that
\eqn\dequalssixo{\eqalign{
&{}\D^{(6)}_{[\ell+2;\ell]+}(s,\x)\,\D^{(6)}_{[\ell'+2;\ell']-}(s,\x)\cr
&{}=\A^{(6)}_{[\ell+\ell'+4;(\ell,\ell,\ell)\otimes
(\ell',\ell',-\ell')]}(s,\x)
+\sum_{q=1}^{\infty}
\D^{(6)}_{[\ell+\ell'+q+4;\ell+\ell'+q,\ell+\ell',\ell-\ell']}(s,\x)\,,}
}
and that
\eqn\dequalssixo{\eqalign{
\D^{(6)}_{[\ell+2;\ell]\pm}(s,\x)\,\D^{(6)}_{[\ell'+2;\ell']\pm}(s,\x)
=\sum_{q=0}^{\infty}\sum_{t=|\ell-\ell'|}^{\ell+\ell'}
\D^{(6)}_{[\ell+\ell'+q+4;\ell+\ell'+q,t,\pm t]}(s,\x)\,.}
}
Here and in the following we are using the short-hand notation,
for $r=[{1\over 2}d]$,
\eqn\shorthand{
\chi^{(d)}_{(\ell_1,\dots,\ell_r)\otimes 
(\ell'_1,\dots,\ell'_r)}(\x)\equiv 
\chi^{(d)}_{(\ell_1,\dots,\ell_r)}(\x)
\chi^{(d)}_{(\ell'_1,\dots,\ell'_r)}(\x)\,,
}
in $\A_{[\De;\uell]}(s,\x)$.  Of course
\shorthand\ may be decomposed in terms of $SO(d)$
characters once we know how $\ell\otimes \ell'$
decomposes into irreducible representations.

More generally there is a distinction in such product
formulae between the cases where the dimension is
divisible by four or not so.
 
Explicitly, we have for $d=4m$ that
\eqn\productone{\eqalign{
& \D^{(4m)}_{[\ell+2m-1;\ell]+}(s,\x)\,\D^{(4m)}_{[\ell'+2m-1;\ell']-}(s,\x)
\cr
& =
\sum_{q=0}^{\infty}\sum_{{t_{i}\geq\ell-\ell'\atop
t_i\geq t_{i+1}}}^{\ell+\ell'}
\D^{(4m)}_{[\ell+\ell'+q+4m-2;\ell+\ell'+q,t_1,t_1,t_2,t_2,\dots,
t_{m{-1}},t_{m{-1}},
\ell-\ell']}(s,\x)\,,}
}
and
\eqn\productones{\eqalign{
& \D^{(4m)}_{[\ell+2m-1;\ell]\pm}(s,\x)\,
\D^{(4m)}_{[\ell'+2m-1;\ell']\pm}(s,\x)
\cr
&{}=\A^{(4m)}
_{[\ell+\ell'+4m-2;(\ell,\dots,\pm\ell)
\otimes (\ell',\dots,\pm\ell')]}(s,\x)\cr
& \qquad+
\sum_{q=1}^{\infty}\sum_{{t_{i}\geq|\ell-\ell'|\atop
t_i\geq t_{i+1}}}^{\ell+\ell'}
\D^{(4m)}_{[\ell+\ell'+q+4m-2;\ell+\ell'+q,\ell+\ell',t_1,t_1,t_2,t_2,\dots,
t_{m{-1}},\pm t_{m{-1}}]}(s,\x)\,,}
}
while for $d=4m+2$ we have that
\eqn\productonetp{\eqalign{
&
\D^{(4m+2)}_{[\ell+2m;\ell]+}(s,\x)\,\D^{(4m+2)}_{[\ell'+2m;\ell']-}(s,\x)\cr
&
= \A^{(4m+2)}
_{[\ell+\ell'+4m;(\ell,\dots,\ell)\otimes (\ell',\dots,-\ell')]}(s,\x)\cr
& \qquad + \sum_{q=1}^{\infty}\sum_{t_i\geq \ell-\ell'\atop
t_i\geq t_{i+1}}^{\ell+\ell'}
\D^{(4m+2)}_{[\ell+\ell'+q+4m;\ell+\ell'+q,\ell+\ell',t_1,t_1,t_2,t_2,\dots,
t_{m{-1}},t_{m{-1}},
\ell-\ell']}(s,\x)\,,}
}
and 
\eqn\productonet{
\D^{(4m+2)}_{[\ell+2m;\ell]\pm}(s,\x)\,\D^{(4m+2)}_{[\ell'+2m;\ell']\pm}(s,\x)
=\sum_{q=0}^{\infty}\sum_{t_i\geq |\ell-\ell'|\atop
t_i\geq t_{i+1}}^{\ell+\ell'}
\D^{(4m+2)}_{[\ell+\ell'+q+4m;\ell+\ell'+q,t_1,t_1,t_2,t_2,\dots,
t_{m},\pm t_{m}]}(s,\x)\,.}

A special case of the
previous is the product involving the character
$\D^{(2r)}_{[r-1;0]}$ in \defsr,
corresponding
to a free scalar field, for which we have
\eqn\charfeeeven{
\D^{(2r)}_{[r-1;0]}(s,\x)\,\D^{(2r)}_{[\ell+r-1;\ell]\pm}(s,\x)
=
\sum_{q=0}^{\infty}\D^{(2r)}_{[\ell+q+2r-2;\ell+q,\ell,\dots,\pm\ell]}(s,\x)
\,.
}

Another special case of the above contains a result first found by
Vasiliev \Vas\ which 
generalises a well known result by Flato and Fronsdal
\Flato\ in three dimensions to even dimensions $d=2r$.
This result involves products of the
representation corresponding to the free Dirac spinor,
\eqn\charfanot{\eqalign{
{\it Di}^{(2r)}(s,\x)&{} \equiv\D^{(2r)}_{[r-{1\over2},{1\over 2}]+}(s,\x)+
\D^{(2r)}_{[r-{1\over2},{1\over 2}]-}(s,\x)\cr
& {} =s^{r-{1\over 2}}(1-s)\big(\chi^{(2r)}_{({1\over 2},\dots,{1\over
2},{1\over 2})}(\x)+\chi^{(2r)}_{({1\over 2},\dots,{1\over
2},-{1\over 2})}(\x)\big)P(s,\x)\cr
&{} = s^{r-{1\over 2}}(1-s)(x_1{}^{1\over 2}+x_1{}^{-{1\over 2}})\dots
(x_r{}^{1\over 2}+x_r{}^{-{1\over 2}})P(s,\x)\,.}
}
Using the above product formulae we may show that
\eqn\prodvasiliev{\eqalign{
&{}{\it Di}^{(2r)}(s,\x){\it Di}^{(2r)}(s,\x)\cr
&{}=2
\A^{(2r)}_{[2r-1,0,\dots,0]}\cr
&{} \quad 
+2 \sum_{q=0}^{\infty}\Big(\D^{(2r)}_{[2r+q-1,q+1,1,1,\dots,1,0]}(s,\x)+
\D_{[2r+q-1,q+1,1,1,\dots,1,0,0]}(s,\x)\cr
&{} \qquad \quad \qquad + \dots
+\D^{(2r)}_{[2r+q-1,q+1,1,0,\dots,0,0]}(s,\x)
+\D^{(2r)}_{[2r+q-1,q+1,0,0,\dots,0,0]}(s,\x)\Big) \cr
&{} \quad +\sum_{q=0}^{\infty}\big(\D^{(2r)}_{[2r+q-1,q+1,11,\dots,1,1]}(s,\x)+
\D^{(2r)}_{[2r+q-1,q+1,1,1,\dots, 1,-1]}(s,\x)\big)\,,}
}
regardless of whether $d=2r$ is divisible by four or not so.
The latter  matches Vasiliev's result.

\noindent{\bf Product formulae: odd dimensions}

For $SO(2r+1,2)$ we have that
\eqn\findthatbk{
P^{(2r+1)}(s,\x)
=\sum_{p,q=0}^{\infty}s^{2p+q}\chi^{(2r+1)}_{(q,0,\dots,0)}(\x)\,,
}
where
\eqn\findthatck{\eqalign{
& \chi^{(2r+1)}_{(q,0,\dots,0)}(\x)
=\half {\rm det}[x_i{}^{k_j}-x_i{}^{-k_j}]
\De(x_1+x_1{}^{-1},\dots,x_r+x_r{}^{-1})^{-1}\cr
&{} \times(x_1{}^{1\over
2}-x_1{}^{-{1\over 2}})^{-1}\dots (x_r{}^{1\over
2}-x_r{}^{-{1\over 2}})^{-1}\,,}
}
with $k_1=q+\half +r-1,\,k_j=\half +r-j,\,j>1$.

From the results of
appendix D, we have the following expansions for the free spinor case
of \defto\ and the free scalar case of \deft, namely,
\eqn\expoddo{
\D^{(2r+1)}_{[r;{1\over
2}]}(s,\x)=\sum_{q=0}^{\infty}s^{r+q}\chi^{(2r+1)}_{({1\over
2}+q,{1\over 2},\dots,{1\over 2})}(\x)\,,
}
for the free spinor case
and 
\eqn\expoddoro{
\D^{(2r+1)}_{[r-{1\over 2};0]}(s,\x)
=\sum_{q=0}^{\infty}s^{r+q-{1\over 2}}\chi^{(2r+1)}_{(q,0,\dots,0)}(\x)\,,
}
for the free scalar case.

For \defro\ and $p=r-j$ we have that, for $\ell>\ell_1>\dots >\ell_j$,
\eqn\csixteenanalopodd{\eqalign{
& \D^{(2r+1)}_{[\ell+r+j;\ell,\ell_1,\dots,\ell_j]}(s,\x)\cr
&{} =\!\!\!\sum_{p_1,\dots,p_j,q,t=0}^{\infty}
\sum_{i_1=-{1\over 2}p_1}^{{1\over 2} p_1}\!\dots
\!\sum_{i_j=-{1\over 2}p_j}^{{1\over 2} p_j}\!\!\!
s^{\ell+r+j+q+t+p_1+\dots+ p_j}
\chi^{(2r+1)}_{(\ell+q,\ell,\dots,\ell,\ell_1+2i_1,\dots,\ell_j+2 i_j)}(\x)\,,}
}
which we show in appendix D.
Again, the weights
$(\ell+q,\ell,\dots,\ell,\ell_1+2i_1,\dots,\ell_j+2 i_j)$
may lie outside the dominant Weyl chamber, i.e. not satisfy
\piffles,  for particular 
$\ell,\ell_i$.   For such weights we may use that
$\chi^{(2r+1)}_{\uell'}(\x)={\rm sgn}(w)\chi^{(2r+1)}_{\uell'^{w}}(\x)$,
for some $w\in \W_{2r+1}$,
to relate such characters to those with
dominant integral highest weights $\uell'^{w}$.

Just as for even 
dimensions in \anothersimp\
a simplification to \csixteenanalopodd\ occurs for the
$p=1$ case of \defro\ for $\ell_1\equiv \ell,\,\ell_2=\dots=\ell_r=0$
for which, 
\eqn\anothersimpodd{
\D^{(2r+1)}_{[\ell+2r-1;\ell,0,\dots,0]}(s,\x)=\sum_{p,q=0}^{\infty}
\sum_{k=0}^{\ell}s^{\ell+2r+2p+q-1}
\chi^{(2r+1)}_{(q+k,\ell-k,0,\dots,0)}(\x)\,.
}

Regarding products of free representations,
we may determine that, using \defro\ for $p=1$,
\eqn\prododdo{\eqalign{
&\D^{(2r+1)}_{[r;{1\over
2}]}(s,\x)\,\D^{(2r+1)}_{[r;{1\over
2}]}(s,\x)= \A_{[2r;0,\dots,0]}(s,\x)  \cr
& + \sum_{q=0}^{\infty}\big(
\D^{(2r+1)}_{[2r+q;q+1,1,\dots,1]}(s,\x)+
\D^{(2r+1)}_{[2r+q;q+1,1,\dots,1,0]}(s,\x)+\dots +
\D^{(2r+1)}_{[2r+q;q+1,0,\dots,0]}(s,\x)\big)\,,}
}
and 
\eqn\prododdo{
\D^{(2r+1)}_{[r;{1\over
2}]}(s,\x)\,\D^{(2r+1)}_{[r-{1\over 2};0]}(s,\x)
= \sum_{q=0}^{\infty}
\D^{(2r+1)}_{[2r+q-{1\over 2}
;q+{1\over 2},{1\over 2},\dots,{1\over 2}]}(s,\x)\,,
}
along with
\eqn\prododdo{
\D^{(2r+1)}_{[r-{1\over 2};0]}(s,\x)\,\D^{(2r+1)}_{[r-{1\over 2};0]}(s,\x)
= \sum_{q=0}^{\infty}
\D^{(2r+1)}_{[2r+q-1
;q,0,\dots,0]}(s,\x)\,,
}
which generalise similar formulae obtained in \Flato\ to odd dimensions.

\newsec{Partition functions}

As a partial check of the character formulae corresponding
to free fields, we will compare them to partition functions
which have been obtained by various authors in conformally
invariant theories on $S^1\times S^{d-1}$.  For
these cases, the single
particle partition
function for a local free operator $F$ may expressed by
\eqn\partitionfns{
Y^{(d)}_F(s)=\sum_{q=0}^{\infty}\! n^{(d)}_{F,q}\,s^{\De_0+q}\,,
}
where $n^{(d)}_{F,q}$ enumerates the descendants 
of $F$ in  the flat background $\Bbb{R}^d$.
For even $d$, the form \csixteenanal\ for the character formula
corresponding to
such free fields allows us to obtain $Y^{(d)}_F(s)$ directly
when we set $x_1,\dots,x_{{1\over 2}d}=1$.  Thus we easily find
that
\eqn\partitionfns{
Y^{(d)}_{F_{\pm}}(s)
=\sum_{q=0}^{\infty}n^{(d)}_{F_{\pm},q}\,\,
s^{\ell+{1\over 2}d+q-1}\,,
}
where
for $\ell\neq 0$ (from appendix B)
\eqn\partitionnos{\eqalign{
n^{(d)}_{F_{\pm},q}&{}=
\chi^{(d)}_{(\ell+q,\ell,\dots,\ell,\pm\ell)}(1,\dots,1)=
{\rm dim}(\I^{(d)}_{(\ell+q,\ell,\dots,\ell)})\cr
&{}=2^{{1\over 2}d-1}\prod_{i=1}^{{1\over 2}d-1}
{1\over (d-2i)!}\,(q+i)(2\ell+q+d-2-i)
\prod_{2\leq k<j\leq {{1\over 2}d}}\!\!\!\!(j-k)(2\ell+d-j-k)\,,
}}
while for the scalar field
\eqn\partitionnosp{\eqalign{
n^{(d)}_{S,q}&{}=\chi^{(d)}_{(q,0,\dots,0)}(1,\dots,1)\cr
&{}={2q+d-2\over q+d-2}\left({q+d-2\atop q}\right)\,,
}}
which is the dimension of the rank-$q$ symmetric traceless
tensor representation
of $SO(d)$.  (This agrees with a similar formula in \cardy.)
For chiral Weyl fermions we find, from \partitionnos\
for $\ell={1\over 2}$,
\eqn\partitionnosq{
n^{(d)}_{f_{\pm},q}
=2^{{1\over 2}d}{1\over q!}(q+1)(q+2)\dots(q+d-2)\,.
}
Similarly for the ${1\over 2} d$-form field strength, from \partitionnos\
for $\ell=1$,
\eqn\partitionnosr{
n^{(d)}_{V_{\pm},q}={d\over 2(2q+d)}\left({d\atop {1\over 2}d}\right)
\left({q+d-1\atop q}\right)\,.
}
Note that it may be easily checked that these occupancy
numbers
agree
with those obtained in \kut. for $d=4,6$
(where for $d=4$ then $Y^{(4)}_{V^{+}}(s)+Y^{(4)}_{V^-}(s)$
is the single particle partition function of
the Maxwell field).

For bosonic $F$ the multi-particle partition function is given by
\eqn\multipart{
Z^{(d)}_F(s)={\rm exp}\Big(\sum_{n=1}^{\infty}{1\over
n}
Y^{(d)}_F(s^n)\Big)=\prod_{q=0}^{\infty}(1-s^{\De_0+q})^{-n^{(d)}_{F,q}}\,,
}
while for fermionic $F$ it is given by
\eqn\multiparto{
Z^{(d)}_F(s)={\rm exp}\Big(\sum_{n=1}^{\infty}{1\over
n}(-1)^{n+1}
Y^{(d)}_F(s^n)\Big)=\prod_{q=0}^{\infty}(1+s^{\De_0+q})^{n^{(d)}_{F,q}}.
}
It is easy to check that \multipart, \multiparto\
for the scalar, Weyl fermion
and field strength cases above match the results of \kut\
for $d=4,6$.

Performing the summation in \partitionfns\ for the scalar, Weyl fermion
and ${1\over 2}d$-form field-strength cases we find
\eqn\resum{\eqalign{
Y^{(d)}_S(s)&{} =(s+1)\,{s^{{1\over 2}d-1}\over (1-s)^{d-1}}\,,\cr
Y^{(d)}_{f_{\pm}}(s)& {} =2^{{1\over 2}d}\,{ s^{{1\over 2}(d-1)}
\over (1-s)^{d-1}}\,,\cr
Y^{(d)}_{V_{\pm}}(s)& {} ={d!\over
2 \,({1\over 2}d)!^2}\,{s^{{1\over 2}d}\over (1-s)^{d-1}}
F(1,-\half d+1;\half d+1;s)\,,\cr}
}
(where $F(a,b;c;x)$ is the usual hypergeometric function)
which may be read off directly from \defs, \defsr\ for $x_i=1$.
This form may be directly compared to similar results in
\aha\
whereby the formulae agree for the scalar and Weyl fermion cases.

  It is not difficult to compute the first
couple of numbers \partitionnosr\ for the self-dual $r={1\over 2} d$ form
field strength which we denote by
$F_{\mu_1\dots\mu_{r}}={}^{*}F_{\mu_1\dots\mu_{r}}$.
For $q=0$ this number just counts the number of independent
components in $F_{\mu_1\dots\mu_{r}}$.
Anti-symmetry in the indices implies $\left(2r \atop r\right)$
independent components which is reduced by a factor of
a half due to self-duality.  For $q=1$ \partitionnosr\ counts the
number of first-order descendants, $\pr_\mu F_{\mu_1\dots\mu_{r}}$.
This will be $r \left({2r\atop r}\right)$ less the
number of constraint equations $\pr^{\mu_1}F_{\mu_1\dots\mu_{r}}=0$
which is $\left({2r\atop r-1}\right)$.

As a further example and check of our formulae, we consider the single 
particle partition function for rank-$\ell$
symmetric
traceless
tensor fields $T_{\mu_1\dots \mu_\ell}$ satisfying
the constraint equation $\pr^{\mu_1}T_{\mu_1\dots \mu_\ell}=0$.
The appropriate character formula in this case is
given by \anothersimp.
The corresponding occupation numbers,
for  $n^{(d)}_{F,q}\to N_{q,\ell}$ in
\partitionfns, are given by 
\eqn\simpn{\eqalign{
 N_{q,\ell}=&
\Big(q(d-2)(2\ell+d-3)+(d-1)(\ell+d-3)(2\ell+d-2)\Big)\cr
&\,\,\times{1\over (d-1)(d-2)(d-3)} \left({\ell+d-4\atop \ell}\right)
\left({q+d-2\atop q}\right)\,.}
}
To see this we may use \anothersimp\ to write 
\eqn\simpextra{
N_{q,\ell}=\sum_{i=0}^{[{1\over 2}q]}t_{q-2 i,\ell}\,\,\,\,{\rm
for}\,\,\,\,
t_{q,\ell}=\sum_{k=0}^{\ell}{\rm
dim}(\I^{(d)}_{(q+k,\ell-k,0,\dots,0)})\,.
}
Using  the dimension formula in appendix B we may find that
\eqn\simpleocc{\eqalign{
t_{q,\ell}
& =(\ell+q+d-3)
\Big(4\ell q+(d-3)(2\ell+2q+d-2)\Big)\cr
& \,\,\,\,\,\,\times {1\over (d-2)(d-3)^2}\left({\ell+d-4\atop \ell}\right)
\left({q+d-4\atop q}\right)\,,}
}
and thence obtain \simpn\ from \simpextra.

Again it is not difficult to check that the first couple
of numbers agree with expectations.  We may easily show that $N_{\ell,0}=
{\rm dim}(\I^{(d)}_{(\ell,0,\dots,0)})$ given in \partitionnosp\
as expected.  Also we may easily show that
$N_{\ell,1}=d\,{\rm dim}(\I^{(d)}_{(\ell,0,\dots,0)})-{\rm
dim}(\I^{(d)}_{(\ell-1,0,\dots,0)})$
which is the number of first-order descendants $\pr_\mu
T_{\mu_1\dots\mu_\ell}$
reduced by the number of constraint equations coming from the
conservation condition.  More
generally
\eqn\gytop{
N_{\ell,q}=\left({q+d-1\atop q}\right){\rm
dim}(\I^{(d)}_{(\ell,0,\dots,0)})
-\left({q+d-2\atop q-1}\right){\rm
dim}(\I^{(d)}_{(\ell-1,0,\dots,0)})\,,
}
which may be easily seen as the descendants at level
$q$ are given by $\pr_{\nu_1}\pr_{\nu_2}\dots \pr_{\nu_q}
T_{\mu_1\dots\mu_\ell}$ whose number of independent components
is given by the first term in \gytop\ which is
reduced by the number of independent components
in $\pr_{\nu_1}\pr_{\nu_2}\dots \pr_{\nu_{q-1}}\pr^{\mu_1}
T_{\mu_1\dots\mu_\ell}$ which vanishes by conservation. 

More generally, we may use \csixteenanalopo\ in even dimensions and
\expoddo,
\expoddoro, \csixteenanalopodd\ in odd dimensions
for $x_i=1$ to determine the occupation numbers in the single particle 
partition function
corresponding to fields whose conformal dimension
saturates the unitarity bounds 
\unitarityd, \unitarityodd. As mentioned before these are fields which
satisfy certain conservation conditions which determines 
the particular unitarity bound.

Rotating quantum fields in an AdS$_{d+1}$ background
have been considered in \gib.  Here the modes
of a quantum field are supposed to have energies $E$ and angular momenta
$j_i$ where $i=1,\dots,[{1\over 2}d]$.  For the boundary conformal
field theory on
$\Bbb{R}\times S^{d-1}$ the energies are related to the
conformal dimension of conformal
fields and their descendants via $E=\De$
assuming that the sphere has unit radius
while $j_i$ correspond to $SO(d)$ eigenvalues.
Making the identification
\eqn\varspartgib{
s=e^{-\be}\,,\qquad x_i=e^{\be\Omega_i}\,,
}
where $\be=T^{-1}$ for $T$ being
the temperature and $\Omega_i$ denote chemical
potentials for angular momenta, then it can be
shown that the one particle boundary partition function
$\sum_{E,j_i}e^{-\be (E-\Omega_i j_i)}$ 
and character
formula for the conformal field coincide.
For instance, for a free scalar field, the character
formula \defsr\ obtained here agrees with the corresponding single
particle partition function for the boundary
conformal field theory obtained in \gib\ when
we make the identification \varspartgib.

%\newsec{Conclusion}

%During the course of this work we noticed that
%a formula which gave the same result for the
%character of the $SO^*(d+2)$ representations
%considered here as that explained in appendix
%A was simply
%\eqn\charnondom{
%\chi_{\Lambda}=\C_{\Lambda}+\sum_{w\in\W\atop
%\Lambda^w\prec \Lambda}{\rm sgn}(w)\C_{\Lambda^w}\,.
%}
%It would be interesting to know to what extent
%this formula is true for the character of
%irreducible representations of semi-simple
%Lie algebras having highest weights which are not
%dominant integral.
%Despite our stealthiest efforts we have failed
%to find a reference to such a formula in the
%literature.
%Assuming this formula leads to a much more
%succinct argument for what the characters of
%the
%positive energy unitary irreducible representations
%are.   The arguments of \gru\ and the
%extensions given here come close to this
%formula. A proof of \charnondom\ obviously 
%requires knowing the extent to which we can 
%further determine $w$ satisfying condition (A.9)
%and give
%a more general formula for $\gamma_{\Lambda^w}$.
%Notice that in \charnondom\ condition
%(A.9) is not assumed to hold - merely the
%weaker condition that $\Lambda^w\prec \Lambda$.
%We are aware of a formula in \enr\ which gives
%the character of unitary highest weight modules
%but at least for the unitary highest weight
%irreducible modules of $SO(d,2)$ then \charnondom\
%seems a much simpler formula to use, were it
%to be true, of course.

\newsec{Acknowledgements}
I am grateful to Nicolas Boulanger,
Gary Gibbons, Paul Heslop and Hugh Osborn
for discussions.

\appendix{A}{Character formulae for infinite dimensional
irreducible modules of semi-simple Lie algebras}

In this appendix we outline some
results on character formulae
of relevance for the main text.
First we give basic definitions and notation.

The Weyl group $\W$ is generated by mappings $w_\al$,
for $\al$ being a root, which
on weight space give
\eqn\first{
w_{\al}(\Lambda)
=\Lambda-(\al^\vee,\Lambda)\,\al\,,\qquad \al^\vee=2\al/(\al,\al)\,.
}
For $\Lambda=\be$, another root, then $w_\al(\be)$ is
a reflection of $\be$ with respect to the hyperplane
through the origin and perpendicular to $\al$.
Here $(\lambda,\mu)$ denotes the usual
inner product on weight space between $\lambda$ and $\mu$.
(In the Dynkin basis this is given by
$(\lambda,\mu)=\sum_{i,j}\lambda_i G_{ij}\mu_j$
where $\lambda_i,\,\mu_i$ are Dynkin labels and $[G_{ij}]$
is the quadratic form matrix.)
%The mapping \first\ is a Weyl reflection and 
%the set of all such reflections forms a group called
%the Weyl group. We will
%denote this by $\W$.
%$\W$ group has
%generators given by simple Weyl reflections,
%$w_{\al_i}\equiv w_i$ for $\al_i$ being a simple root.
Any %other element of the Weyl group 
$w\in \W$  may be decomposed
in terms of simple Weyl reflections $w_i\equiv w_{\al_i}$,
for $\al_i$ being the simple roots,
as
$w=w_{i_1}\cdots w_{i_n}$
for some $n$ which is generally not unique.
However the signature of $w$
defined, in the present case, by
${\rm sgn}(w)=(-1)^n$
is uniquely defined.
We denote by $\ell(w)$ the minimum number of $w_i$
in the composition of $w$.  Clearly ${\rm sgn}(w)=(-1)^{\ell(w)}$.

%On simple roots themselves we have from \first\ that
%\eqn\firstt{
%w_i(\al_j)=\al_j-K_{ij}\al_i\,,
%}
%for $[K_{ij}]$ being the Cartan matrix.
%A number of important points follow from \firstt.
%First it implies that $w_i$ and $w_j$ commute
%for $K_{ij}=0$.  Also, $w_i(\al_i)=-\al_i$
%and since $K_{ij}\leq 0$ for $i\neq j$ then $w_i$ permutes all other positive
%roots so that $w_i\big(\Phi_+/\{\al_i\}\big)=\Phi_+/\{\al_i\}$,
%where $\Phi_+$ denotes the set of positive roots.
%From here it is not difficult to show that
%the Weyl vector $\rho={1\over 2}\sum_{\al\in \Phi_+}\al$ has Dynkin
%labels given by $\rho_i\equiv(\al_i{}^{\vee},\rho)=1$.
%The action of the Weyl group thus far defined, extends naturally
%to any weight system.  For an arbitrary weight $\lambda$ we define
%$w_{\al}(\lambda)=\lambda-(\al^{\vee},\lambda)\,\al$.
%Thus defined, Weyl reflections leave invariant the inner product on
%weight space
%$(\lambda,\mu)=(w_\al(\lambda),w_\al(\mu))$
%and so any Weyl group element is also an isometry.

The Weyl group divides the weight space into a family
of open sets called Weyl
chambers.  These are simplicial cones
defined by
\eqn\sixtho{
\H_w=\{\lambda : (\al_i{}^{\vee},w(\lambda))> 0\,,\quad 1\leq i\leq r\}\,,
}
for $w\in\W$.  The number of such equals the order of $\W$, $|\W|$.
The weights lying on the boundary of the Weyl chambers
are the points on the hyperplanes perpendicular to
the roots $(\al_i{}^{\vee},w(\lambda))=0$.  In terms of
Dynkin labels these are the weights having at least one
vanishing Dynkin label.  

The Weyl chamber corresponding to the identity of the Weyl
group $\H_{1}$ is the fundamental or
dominant Weyl chamber.  In terms of Dynkin labels
the weights in this chamber have strictly
positive Dynkin labels.  If all the Dynkin labels
are non-negative and/or integers then we say the weight
is dominant and/or integral.

Suppose we have some Lie algebra module ${V}_{\Lambda}$
having  highest weight $\Lambda$.  The definition
of the corresponding character we use
is \fuchs\
\eqn\chardefo{
{\rm Char}_{\Lambda}\equiv \sum_{\lambda\in V_\Lambda}{\rm
  mult}_{V_\Lambda}(\lambda)\,e^{\lambda}\,,
}
where ${\rm
  mult}_{V_\Lambda}(\lambda)$, ${\rm
  mult}_{V_\Lambda}(\Lambda)=1$,
 denotes the multiplicity
of the weight $\lambda$ in the weight system of $V_{\Lambda}$.
This is to be interpreted as a function on weight space
satisfying
\eqn\fucntwe{
e^{\lambda}e^{\mu}=e^{\lambda+\mu}\,,\qquad e^{\lambda}(\mu)=
e^{(\lambda,\mu)}\,.
}
Under the action of the Weyl group, for $w\in \W$,
\eqn\weyl{
w(e^\lambda)=e^{w(\lambda)}\,.
}

For a unitary group and with ${\rm
  mult}_{V_\Lambda}(\lambda)$ always finite
we may recover a trace formula for the character
by normalising each vector $v_\lambda\in {V}_{\Lambda}$
corresponding to the weight $\lambda$
so that
$\l v_\lambda|v_\lambda\r =1$.
Then we may write,
\eqn\chardefop{
{\rm Char}_{\Lambda}(\mu)=
\sum_{v_\lambda\in {V}_\Lambda}\l v_\lambda|v_\lambda\r
\,e^{(\lambda,\mu)}={\rm Tr}\big(e^{(H,\mu)}\big)\,,
}
where $(H,\mu)=\sum_{i,j}H_iG_{ij}\mu_j$ in the Dynkin basis,
for example, for $H_i$ being Cartan subalgebra elements with
$H_i|v_{\lambda}\r=\lambda_i|v_\lambda\r$.

As an example, consider a Verma module ${\V}_{\Lambda}$ with basis
$\prod_{\al\in\Phi_-}E_{\al}{}^{n_{\al}}|{\underline \Lambda}\r^{\rhw}$
for $\Phi_-$ being negative roots and $n_{\al}$ being
non-negative integers.  For fixed $n_{\al}$ then the corresponding
weight $\lambda_{(n_\al)}=
\Lambda+\sum_{\al\in\Phi_- }n_\al \al$ has
unit multiplicity in the weight system of $\V_{\Lambda}$.
Thus we may write the character for the Verma module as
\eqn\cozero{
\C_\Lambda=\sum_{n_\al\Geq 0}e^{\lambda_{(n_\al)}}=
e^\Lambda\prod_{\al\in\Phi_-}(1-e^{\al})^{-1}\,.
}
Note that a given weight $\lambda$
has multiplicity given by $\P(\Lambda-\lambda)$
where $\P(\mu)$ counts the number of ways in which the
weight $\mu$ may be written as a linear combination of 
positive roots with non-negative integer coefficients.

We may easily also show that
\eqn\formulasym{
w(\C_{\Lambda})={\rm sgn}(w)\C_{\Lambda^w}\,,\qquad 
\Lambda^w=w(\Lambda+\rho)-\rho\,,
}
for any $w\in \W$.

The character $\chi_\Lambda$
of an infinite dimensional irreducible module $\I_\Lambda$ of a 
semi-simple Lie algebra has been written down long ago \gru.
For the highest weight $\Lambda$ not being dominant integral
then $\I_\Lambda$ is infinite dimensional.  Otherwise
$\I_\Lambda$ is finite dimensional and the character is given by
the well known Weyl character formula. 
Before we give the result of \gru\ we quote
a number of results which give insight into the
structure of infinite dimensional irreducible
modules.

Concerning Verma modules (which in \gru\ are called
`elementary representations'),   
the first result we recount is that if 
$\Lambda^w,\,w\in \W$ is not a weight of $\V_\Lambda$ then
$\V_\Lambda$ itself is irreducible.  This is the simplest
case and an example is $\V_\ell$ for $SO(3)$ with $\ell$ 
being a negative half integer.
We now give the conditions for
$\V_{\Lambda}$ 
to contain sub-modules $\V_{\Lambda'}$
and so to be reducible.

We define a partial ordering on
weights so that $\Lambda'\prec\Lambda$ if and only
if $\Lambda-\Lambda'=\Pi $ for some $\Pi=\sum_{\al\in\Phi_+}p_\al \al$
for $p_\al$ being non-negative integers,
not all zero.
A necessary condition for a Verma module
$\V_\Lambda $ to contain a sub-module $\V_{\Lambda'}$
is that $\Lambda'=\Lambda^w\prec \Lambda$
for some $w\in \W$,
$w\neq 1$.  
A crucial result is a  theorem in \ber\
which proves that a necessary and sufficient
condition for $\V_{\Lambda^w}$ to be a sub-module
of  $\V_{\Lambda}$ is that
there exist a sequence of positive roots
$\be_1,\dots,\be_K$ such that 
\eqn\conda{
w=w_{\be_1}w_{\be_2}\dots w_{\be_K}\,,\quad 
\big(w_{\be_k}{}^\vee,w_{\be_{k+1}}w_{\be_{k+2}}\dots 
w_{\be_K}(\Lambda+\rho)\big)\in \Bbb{N}\,,
\quad k=1,\dots,K\,,
}
where we define $w_{\be_{K+1}}\equiv 1$.
For $\Lambda$ being a dominant integral
weight then condition \conda\ holds for all
$\Lambda^w,\,w\in \W$.
This justifies the claim made in this paper that
for $\Lambda$ being dominant integral then $\Lambda^w$
is a highest weight in $\V_\Lambda$ for every $w\in\W$.

A key result proved by Verma \ver\ and recounted in
\gru\ is the following.  A Verma module contains 
those and only those irreducible representations
$\I_{\Lambda'}$ for which $\V_{\Lambda'}$ is a 
sub-module of $\V_\Lambda$.  Furthermore it contains
$\I_{\Lambda'}$ at most once.  

This results in the following formula for characters,
\eqn\charform{
\C_{\Lambda}=\chi_{\Lambda}+\sum_{{w\in\W
\atop \Lambda^{w}\prec \Lambda}}\chi_{\Lambda^{w}}\,,
}
where the sum runs over all $w$ for which
\conda\ holds.

Using these results and formulae for multiplicities of weights
determined for Verma modules in terms of the function $\P(\mu)$
above a formula has been given for $\chi_\Lambda$
in \gru.  This may be rewritten in the equivalent form
\eqn\formulachar{
\chi_{\Lambda}=\C_{\Lambda}+\sum_{{w\in \W\atop \Lambda^w\prec
\Lambda}}\gamma_{\Lambda^w}\,
\C_{\Lambda^w}\,,
}
where each $w$ satisfies condition \conda\ and
where the integers $\gamma_{\Lambda^w}$ are determined by a recurrence
relation as follows.

Consider a sub-module $\V_{\Lambda^w}$
of $\V_{\Lambda}$
so that there is no other sub-module
$\V_{\Lambda^{w'}}$
of $\V_{\Lambda}$ containing 
$\V_{\Lambda^w}$ in turn as a sub-module.  Then in this case
$\gamma_{\Lambda^w}=-1$.

For a sub-module $\V_{\Lambda^w}$ of $\V_{\Lambda}$
which is in turn contained in the sub-modules
$\V_{\Lambda^{w'}}$
of $\V_{\Lambda}$, then in this case
\eqn\gammadef{
\gamma_{\Lambda^w}=-\sum_{w'}\gamma_{\Lambda^{w'}}-1\,,
}
which determines $\gamma_{\Lambda^w}$ in \formulachar\ recursively.

A simple example illustrates this formula.
Consider $Sl_3$ whereby, for a weight $\Lambda=[a,b]$
with Dynkin labels $a,b$,
the simple Weyl reflections are
given by
\eqn\simpweylsltw{
w_1([a,b])=[-a,a+b]\,,\qquad w_2([a,b])=[a+b,-b]\,.
}
The $\S_3$ Weyl group elements consist of $\{1,w_1,w_2,w_1w_2,w_2w_1,w_\al\}$
where $w_\al=w_1w_2w_1=w_2w_1w_2$ is the Weyl reflection
corresponding to the root $\al=\al_1+\al_2$.
The shifted Weyl reflections are given by
$
\Lambda^1=\Lambda,\,\Lambda^{w_1}=\Lambda-(a+1)\al_1,\,
\Lambda^{w_2}=\Lambda-(b+1)\al_2,\,\Lambda^{w_2w_1}=\Lambda-(a+1)\al_1
-(a+b+2)\al_2,\,\Lambda^{w_1w_2}=\Lambda-(a+b+2)\al_1-(b+1)\al_2,\,
\Lambda^{w_\al}=\Lambda-(a+b+2)\al
$.
For $\Bsw$ denoting a vector in the $-\al_1$
direction of length $a+1$ and $\Bse$ denoting such
in the $-\al_2$ direction of length $b+1$
then we may represent the weight system 
of the Verma module $\V_\Lambda$ diagrammatically 
as follows
\eqn\rurakanalazare{\def\normalbaselines{\baselineskip16pt\lineskip3pt
\lineskiplimit3pt}
\matrix{
&&{~~~}&&{~~}&& \Lambda  &&{~~}&&{~~~}&\cr
&&&&&\Bsw&&\Bse&&&&\cr
&&&&
\hidewidth~~~~~\Lambda^{w_1}
\hidewidth&&&&
\hidewidth~~ \Lambda^{w_2}~~\hidewidth&&&\cr
&&&&&\Bse&&\Bsw&&&&\cr
&&&&&&&&&&&\cr
&&&&&\Bsw&&\Bse&&&&\cr
&&&
&\hidewidth~~\Lambda^{w_1w_2}\hidewidth&&&
&\hidewidth\Lambda^{w_2w_1}\hidewidth&&&\cr
&&&&&\Bse&&\Bsw&&&&\cr
&&&&&
&\hidewidth \Lambda^{w_\al}\hidewidth&&&&&\cr}
}
assuming $\Lambda$ is dominant integral and
where we have omitted any other weights occurring.
In computing $\chi_{\Lambda^{w_1}}$,
for example, we have that $\V_{\Lambda^{w_1w_2}}$
and $\V_{\Lambda^{w_2w_1}}$ are both sub-Verma-modules of
$\V_{\Lambda^{w_1}}$ with $\gamma_{\Lambda^{w_1w_2}}=
\gamma_{\Lambda^{w_2w_1}}=-1$.
Also $\V_{\Lambda^{w_\al}}$ is a sub-Verma-module of
both the latter and so 
$\gamma_{\Lambda^{w_\al}}=-(\gamma_{\Lambda^{w_1w_2}}
+\gamma_{\Lambda^{w_2w_1}})-1=1$.
Thus $\chi_{\Lambda^{w_1}}=\C_{\Lambda^{w_1}}-
\C_{\Lambda^{w_2w_1}}-\C_{\Lambda^{w_1w_2}}+\C_{\Lambda^{w_\al}}$.
In fact for all $\Lambda'=\Lambda^{w'}$ it is easy to check that
$\chi_{\Lambda'}=\C_{\Lambda'}+
\sum_{{w\in \S_3\atop \Lambda'{}^{w}\prec \Lambda'}}
{\rm sgn}(w)
\C_{\Lambda'{}^{w}}$.

Other properties of $\I_\Lambda$
which are useful for what follows concern
symmetry under Weyl group reflections. For all Dynkin labels
$\Lambda_i$ of the weight $\Lambda$
which are non-negative integers
then the sub-group $\W'$ generated by the corresponding
simple Weyl reflections $w_i$ is the maximal
symmetry group of the weight system of $\I_{\Lambda}$.
Furthermore $\V_{\Lambda^{w'}}$ is a sub-module
of $\V_{\Lambda}$ for all $w'\in \W',\,w'\neq 1$
(since all $w'$ satisfy condition \conda).
In terms of characters we have that $w'(\chi_\Lambda)=\chi_\Lambda $
for every $w'\in \W'$.

For $\Lambda$ being dominant integral then
this symmetry group is of course the Weyl group itself
and $\W'=\W$.  In this case we from \formulachar\
that
\eqn\comfour{
\chi_{\Lambda}=\sum_{w\in \W}{\rm sgn}(w)\C_{\Lambda^w}
=\prod_{\alpha\in\Phi_-}(1-e^{\al})^{-1}
\sum_{w\in \W}{\rm sgn}(w)e^{w(\Lambda+\rho)-\rho}\,,
}
the usual Weyl character formula. 
This follows as symmetry under $\W$ determines
$\gamma_{\Lambda^w}={\rm sgn}(w)$ in this case.
Using \formulasym\ 
we may find that the Weyl character
may be rewritten as
\eqn\comthree{
\chi_\Lambda=\sum_{w\in \W}w(\C_\Lambda)\equiv\frak{W}(\C_\Lambda)\,.
}
Due to the invariance of
$\I_\Lambda$ under the action of any $w\in\W$ then the Weyl 
symmetry operator
$\frak{W}$ defined by \comthree\ is obviously linear and idempotent
on the vector space spanned by the characters of the Verma modules, 
$\frak{W}^2=|\W|\frak{W}$.

%Suppose that $\Lambda_i$ are non-negative
%for $i\in I\subseteq \{1,\dots,r\}$ where $r$ is the rank of the Lie
%algebra.
%We
%claim that any other $\Lambda^w,\,w\in \W',\,w\neq 1$
% has $\Lambda^w{}_i$ not all non-negative.

Denoting by $I$ the subset of labels for which $\Lambda_i$, $i\in I$,
are non-negative integers,
supposing we find that every 
sub-Verma-module 
highest weight $\Lambda^{w''},\,w''\in\W$
may be written in the form $\Lambda^{w''}=\Lambda^{w'w},\,w\in
\W,\,w'\in\W'$ for 
$\Lambda^{w}{}_i,\,i\in I$ being non-negative
integers or $\Lambda^{w}{}_i=-1$
for some $i\in I$. 
We claim that the weights $\Lambda^{w''}\prec \Lambda$ are given by
the
disjoint union of the weights $\Lambda^{w'w}$ for every $w'\in\W'$.
For the cases of $\Lambda^w{}_i=-1$
then the $\C_{\Lambda^{w'w}},\,w'\in \W'$
cancel among themselves in $\chi_\Lambda$.

%Using \formulachar\ and since the character $\chi_\Lambda $
%is symmetric with respect to $\W'$ then 
%every weight
%$\Lambda^{w''}\prec\Lambda$
%may be written as $\Lambda^{w''}=\Lambda^{w'w}$ 
%for all $w'\in\W'$ and some $w\in \W$. 

%On the $\W'$ orbit of any $\Lambda+\rho$ 
%for $\Lambda_i\in \Bbb{Z},\,i\in I$, there is a weight
%$w'(\Lambda+\rho)$ for which
%$\Lambda^{w'}{}_i,\,i\in I$ are non-negative integers
%for all $i$ apart from when $\Lambda^{w'}{}_i=-1$
%for some $i$.
%Suppose that this is not the case and in particular 
%that, for $i\in I$, the $i^{\rm th}$ Dynkin label of every $\Lambda^{w'}$ is
%a negative integer $\Lambda^{w'}{}_i=(\al_i{}^\vee,
%w'(\Lambda+\rho))-1<0$. Then we have that 
%$\Lambda^{w_iw}{}_i=(\al_i{}^{\vee},w_iw'(\Lambda+\rho))-1
%=-(\al_i{}^\vee,w'(\Lambda+\rho))-1$
%must be a non-negative integer or $-1$ which is a contradiction.
%Thus at least one Dynkin label of some $\Lambda^{w'}$ is a non-negative
%integer or $-1$.  Assuming that all Dynkin labels but
%one, say the $j^{\rm th}$
%one, of $\Lambda^{w'}$ is a non-negative integer or $-1$
%then $\Lambda^{w_jw'}_{j}$ is a non-negative integer or $-1$ by
%the previous argument.  Also $\Lambda^{w_jw}_i=
%(w_j(\al_i)^\vee,\Lambda^{w'}+\rho)-1$ and since $w_j(\al_i)=\al_i-
%K_{ji}\al_j$ is a positive root 

Under this assumption (which holds in the cases
considered in this paper)
then using \formulachar\ we may write
for the character
\eqn\charformo{
\chi_{\Lambda}=\C_{\Lambda}+\sum_{w'\in\W'\atop
w'\neq 1}\gamma_{\Lambda^{w'}}\C_{\Lambda^{w'}}
+\sum_{w'\in\W'\atop w\in\W,w\neq 1}\gamma_{\Lambda^{w'w}}
\C_{\Lambda^{w'w}}\,,
}
where the sum runs over only those $w\in \W$ which
satisfy \conda\ and for which $\Lambda^w{}_i$, $i\in I$,
is a non-negative integer or $-1$.
Using \formulasym\ and $w'(\chi_{\Lambda})=\chi_\Lambda,\,w'\in\W'$
and the claims proved below we may show
that $\gamma_{\Lambda^{w'}}={\rm
sgn}(w'),\,\gamma_{\Lambda^{w'w}}=\gamma_{\Lambda^w}{\rm sgn}(w')$ and 
that the $\C_{\Lambda^{w'w}}$ for which $\Lambda^w{}_j=-1$
for some $j\in I$ cancel among themselves.
It is then left to determine
the remaining $\gamma_{\Lambda^w}$ by the recurrence relation
mentioned earlier.

We now show that
for some arbitrary weight $\Lambda$
which has $\Lambda_i, \,i\in I$
being  non-negative integers that
this is the unique weight among $\Lambda^{w'},\,w'\in\W'$
having this property.
Clearly under these assumptions
$(\Lambda^{w_i}){}_i=(\al_i{}^\vee,\Lambda^{w_i})
=-\Lambda_i-2$ is negative for $i\in I$.   
Since $\W'$ is generated by all $w_i,\,i\in I$, then we may consider all
$w'=w_{i_1}\dots w_{i_n},\,i_j\in I$ such that $\ell(w')=n$. 
Denoting by $\Phi_+{}'$
all those positive roots formed from linear combinations
of the subset of simple roots $\al_i,\,i\in I$, then we have the result
that $\ell(w_iw')=\ell(w')+1$ if and only if
$w'{}^{-1}(\al_i)\in \Phi_+{}'$.{\foot{
The argument for this is similar to one
given in \hum.}} 
In this case we have that
$(\Lambda^{w_iw'}){}_i=-(\al^\vee,\Lambda+\rho)-1$ where
$\al=w'{}^{-1}(\al_i)\in\Phi_+{}'$.  Since $(\al^\vee,\Lambda+\rho)\Geq
0$
then 
$(\Lambda^{w_iw'}){}_i$ must be negative.  Hence all
$\Lambda^{w'},\,w'\in \W',\,w'\neq 1$,
 have at least one of $\Lambda^{w'}{}_i,\,i\in I$ which is
negative.

If some $\Lambda^{w'w}=\Lambda^{u'u}$
for some $w',u'\in \W'$ and with $\Lambda^{w}{}_i,\,\Lambda^u{}_i\geq
 0,\,i\in I$,
then clearly we may write $\Lambda^u=\Lambda^{u'^{-1}w'w}$ which
contradicts the above
unless $u=w,\,u'=w'$.  Thus any two sets of such  weights
 $\{\Lambda^{w'w},\,w'\in \W'\}$
and $\{\Lambda^{w'u},\,w'\in\W'\}$ for $w\neq u$ are disjoint.

If for some weight 
$\Lambda_j=-1$
for some $j\in I$
then  $\Lambda^{w_j}=\Lambda$
since $(\al_j{}^\vee,\Lambda^w+\rho)=0$.
We claim that the only $\Lambda^{w'}$ for $w'\in \W'$
which equal $\Lambda$ in this case are those for
which $w'$ is composed of the $w_j$.  For simplicity
we consider the case when 
$\Lambda_j=-1$ and
otherwise $\Lambda_i$ is non-negative
for $i\in I$.  In this case if $\Lambda^{w'}=\Lambda$
then clearly $\Lambda^{w_jw'}=\Lambda$ so
 that,
for $\al=w^{-1}(\al_j)$,
 $(\Lambda^{w_jw'}){}_j=-(\al^\vee,\Lambda+\rho)-1=-1$
by assumption.  Thus $(\al^\vee,\Lambda+\rho)=0$
which is only the case if $\al\propto \al_j$.
The only roots with this property are
$\pm \al_i$ so that $w'=1,w_j$.
Also this implies that, for $w',u'\in\W'$,
$\Lambda^{w'}=\Lambda^{u'}$ if and only if $u'=w'$ or $u'=w'w_j$.
The generalisation is clear.

%We next claim that if $\Lambda_j=-1$ for some $i\in I$
%then $\Lambda^{w'}\preccurlyeq \Lambda$. 

\appendix{B}{Weyl character formulae for $SO(d)$}

We now consider $SO(d)$ character formulae.  We define
the variables $x_i=e^{\e_i}(\mu)=e^{\mu_i}$ for some arbitrary
weight $SO(d)$ $\mu=\sum_{i=1}^r \mu_i \e_i$.

For $SO(2r)$ the action of the Weyl group, $\W_{2r}=\S_r \ltimes
\Bbb{Z}^{r-1}$,
on weights in the orthonormal basis is given by $\S_r$
permutations on the labels followed by reflections involving
an even number of sign flips
in the labels.
This means that for $\varrho=\rho_i\dots \rho_j\in \Bbb{Z}^{r-1}$ where 
$\rho_i(\ell_1,\dots,\ell_i,\dots \ell_r)
=(\ell_1,\dots,-\ell_i,\dots \ell_r),\,\rho_i{}^2=1$
then the number of $\rho_i$ in the composition of $\varrho$ is even and
 ${\rm sgn}(\varrho)=1$.
Using that 
\eqn\cozeroo{
\sum_{\si\in\S_r}{\rm sgn}(\si)x_{\si(1)}{}^{\ell_1}\dots 
x_{\si(r)}{}^{\ell_r} = {\rm det}[x_i{}^{\ell_j}]\,,
} 
and the restriction mentioned on $\varrho\in \Bbb{Z}^{r-1}$
then \fult,
\eqn\coaone{
\frak{W}_{2r}\Big(\prod_{i=1}^r\,x_i{}^{\ell_i}\Big)
= \half {\rm{det}}[x_i{}^{\ell_j}+x_i{}^{-\ell_j}]+
\half {\rm{det}}[x_i{}^{\ell_j}-x_i{}^{-\ell_j}]\,,
}
for $\frak{W}_{2r}$ denoting the $SO(2r)$ Weyl symmetry operator.
For
some highest weight $\ell=\sum_{i=1}^r\ell_i\e_i$ the corresponding
Verma module character is given by,
\eqn\coazero{\eqalign{
C^{(2r)}_{\uell}(\x) & {} \equiv  
\C^{(2r)}_{\uell}(\mu)=\prod_{i=1}^r x_i{}^{\ell_i}
\prod_{1\leq j<k\leq r,\vep=\pm}
(1-e^{(\al_{jk,\vep},\mu)})^{-1}\cr
& {}=  \prod_{i=1}^r x_i{}^{\ell_i+r-i}\,\,\De(x_1+x_1{}^{-1},\dots,
x_r+x_r{}^{-1}){}^{-1}
\,,}
}
where $\al_{ij,\pm}=-\e_i\pm\e_{j},\,1\leq i<j\leq r$
are the negative roots 
and $\De(\x)$ is the Vandermonde
determinant \cthirteeno. Using the fact that $\De(x_1+x_1{}^{-1},\dots,
x_r+x_r{}^{-1})$ is left alone by any $\varrho\,\si\in \S_r\ltimes
\Bbb{Z}^{r-1}$ then we have quite simply that the Weyl
character of the irreducible representation 
with dominant integral highest weight $\ell$ (given by
$\chi^{(2r)}_{\uell}(\x)=\frak{W}_{2r}(C^{(2r)}_{\uell}(\x))$) 
reduces to
\eqn\coatwo{
 \chi^{(2r)}_{\uell}(\x)=\half \big({\rm{det}}[x_i{}^{k_j}+x_i{}^{-k_j}]+
{\rm{det}}[x_i{}^{k_j}-x_i{}^{-k_j}]\big)
 \De(x_1+x_1{}^{-1},\dots,
x_r+x_r{}^{-1}){}^{-1}\,,
}
for $k_i=\ell_i+r-i$.
The dimension of the irreducible representation is given by
\eqn\dimev{
{\rm dim}(\I^{(2r)}_{\uell})=2^{r-1}\prod_{i=1}^{r}{1\over
(2r-2i)!}\prod_{1\leq i<j\leq
r}(\ell_i-\ell_j+j-i)(\ell_i+\ell_j+2r-i-j)\,.
}

For $SO(2r+1)$ the action of the Weyl group, $\W_{2r+1}=\S_r \ltimes
\Bbb{Z}^{r}$,
on weights in the orthonormal basis is given by $\S_r$
permutations on the labels followed by reflections involving
any number of sign flips
in the labels.
Using \cozeroo\ we therefore have that \fult,
\eqn\coaonep{
\frak{W}_{2r+1}\Big(\prod_{i=1}^r\,x_i{}^{\ell_i}\Big)
= {\rm{det}}[x_i{}^{\ell_j}-x_i{}^{-\ell_j}]\,,
}
for $\frak{W}_{2r+1}$ denoting the $SO(2r+1)$ Weyl symmetry operator.
This time for
some highest weight $\ell=\sum_{i=1}^r\ell_i\e_i$ the corresponding
Verma module character is given by,
\eqn\coazerop{\eqalign{
&{} C^{(2r+1)}_{\uell}(\x) \equiv  
\C^{(2r+1)}_{\uell}(\mu)=\prod_{i=1}^r x_i{}^{\ell_i}
\prod_{1\leq j<k\leq r,\vep=\pm}
(1-e^{(\al_{jk,\vep},\mu)})^{-1}\prod_{1\leq l\leq
r}(1-e^{(\e_l,\mu)})^{-1}
\cr
& {}=  \prod_{i=1}^r x_i{}^{\ell_i+{1\over 2}+r-i}\,\,\De(x_1+x_1{}^{-1},\dots,
x_r+x_r{}^{-1}){}^{-1}(x_1{}^{1\over 2}-x_1{}^{-{1\over 2}})^{-1}\dots
(x_r{}^{1\over 2}-x_r{}^{-{1\over 2}})^{-1}
\,,}
}
since $-\e_i,\,1\leq i\leq r$ are the weights of the extra negative
roots
in this case.  Using that $\De(x_1+x_1{}^{-1},\dots,
x_r+x_r{}^{-1}){}^{-1}(x_1{}^{1\over 2}-x_1{}^{-{1\over 2}})^{-1}\dots
(x_r{}^{1\over 2}-x_r{}^{-{1\over 2}})^{-1}$ is left alone by any
$\varrho \,\si\in \S_r \ltimes
\Bbb{Z}^{r}$ then the Weyl character of the irreducible representation
with dominant integral highest weight $\ell$ is given by,
using \coaonep\ and \coazerop,
\eqn\chaross{\eqalign{
&\chi^{(2r+1)}_{\uell}(\x)\cr
&{}={\rm{det}}[x_i{}^{k_j}-x_i{}^{-k_j}]
\De(x_1+x_1{}^{-1},\dots,
x_r+x_r{}^{-1}){}^{-1}(x_1{}^{1\over 2}-x_1{}^{-{1\over 2}})^{-1}\dots
(x_r{}^{1\over 2}-x_r{}^{-{1\over 2}})^{-1}
\,,}
}
where $k_i=\ell_i+{1\over 2}+r-i$.
The dimension of the irreducible representation is given by
\eqn\dimodd{\eqalign{
{\rm dim}(\I^{(2r+1)}_{\uell})=\prod_{i=1}^{r}& {1\over
(2r+1-2i)!}(2 \ell_i+2r+1-2i)\cr
&\times \prod_{1\leq i<j\leq
r}(\ell_i-\ell_j+j-i)(\ell_i+\ell_j+2r+1-i-j)\,.}
}

\appendix{C}{Unitarity bounds}
Descendant states have bases, for $p=0,1,\dots$,
\eqn\copo{
\B^{(p)}=
\left\{\prod_{{v=i\vep,0\atop
1\leq i\leq r,\,\vep=\pm}}\P_{v}{}^{n_{v}}|\De;\uell'\r\,,
\quad \sum_{v}n_{v}=p\right\}\,,
}
for $n_{v}$ being positive integers
(with $n_0=0$ for $SO(2r)$) and
$\ell'$ being a weight in the weight system
of the  
module of $SO(d)$ with highest weight $\ell$. For $p=0$
the norms of corresponding states are strictly
positive for $\ell'$ being weights in the $SO(d)$ irreducible
representation with dominant integral highest weight $\ell$
and
$\big\||\De,\ell'\r\big\|^2=\l\De,{\overline{\uell'}}|\De,\uell'\r>0$.

Examining the simplest descendant states with basis $\B^{(1)}$
then these have $SO(d)$ highest weight states 
\eqn\defshd{
\H_d=\{|\De+1;\uell+\uv\r\}\,,
}
where $\v=\vep\e_i$ along with $\v=0$ for $d=2r+1$,
these of course
occurring in the decomposition of the product
between the vector representation and the representation with highest
weight $\ell$ into irreducible representations, 
$\e_1\otimes \ell=\bigoplus_{\v}\ell\oplus \v$.
Remarkably, most of the
restrictions necessary for the states in $\H_d$ to
have positive definite norm are sufficient for the unitarity
constraints
to be satisfied for all descendant states in $\B^{(p)}$ 
- also conjectured in \min.

The simplest states in $\H_d$ may be constructed explicitly and are  
\eqn\coseventeen{\eqalign{
|\De+1;\uell+\ue_1\r&{}=\P_{1+}|\De;{\uell}\r^{\rhw}\,,
\cr
|\De+1;\uell+\ue_2\r&{}=
(-2i(\ell_1-\ell_2)\P_{2+}
+\P_{1+} E_{12}^{-+}) |\De;{\uell}\r^{\rhw} \,,
\cr
|\De+1;\uell+\ue_3\r&{} =\big(-4(\ell_1-\ell_3+1)(\ell_2-\ell_3)\P_{3+}-2i
(\ell_1-\ell_3+1)
\P_{2+}E_{23}^{-+}\cr
& \,\,\,\,+(\ell_2-\ell_3+1)\P_{1+}E_{12}^{-+}E_{23}^{-+}-(\ell_2-\ell_3)
\P_{1+}E_{23}^{-+}E_{12}^{-+}\big)
|\De;{\uell}\r^{\rhw}\,,\cr
}
}
which are all annihilated by $SO(d)$ raising operators.
Using the conformal algebra
and the unitarity conditions,
the norms of these three states are given by,
\eqn\threestates{\eqalign{
\big\| |\De+1;\uell+\ue_1 \r \big\|^2 & {} =
4(\De+\ell_1)\big\| |\De;\uell\r \big\|^2\,,\cr
\big\| |\De+1;\uell+\ue_2 \r \big\|^2 & {} =
16 (\De+\ell_2-1)(\ell_1-\ell_2)(\ell_1-\ell_2+1)
\big\| |\De;\uell\r \big\|^2\,,\cr
\big\| |\De+1;\uell+\ue_3 \r \big\|^2 & {} =
64 (\De+\ell_3-2)(\ell_1-\ell_3+1)(\ell_1-\ell_3+2)\cr
&\qquad \times (\ell_2-\ell_3)
(\ell_2-\ell_3+1)
\big\| |\De;\uell\r \big\|^2\,,}
}
and for these to be strictly positive this places obvious
restrictions on $\De$.
Other examples which may be readily achieved are, for $SO(3,2)$,
\eqn\statesthree{\eqalign{
& |\De+1;\ell+1\r=\P_{1+}|\De;\ell\r^{\rhw}\,,\quad |\De+1;\ell\r
=(-2i\ell\P_0+\P_{1+}E_{1}^{-})|\De;\ell\r^{\rhw}\,,\cr
& |\De+1;\ell-1\r=(2\ell(2\ell-1)\P_{1-}-2i(2\ell-1)\P_0E_{1}^{-}+\P_{1+}
(E_{1}^{-})^2)|\De;\ell\r^{\rhw}\,,}
}
for which the norms are
\eqn\normsthese{\eqalign{
\big\| |\De+1;\ell+1 \r \big\|^2 & {} =
4(\De+\ell)\big\| |\De;\ell\r \big\|^2\,,\quad
\big\| |\De+1;\ell \r \big\|^2 =
8(\De-1)\ell(\ell+1)\big\| |\De;\ell\r \big\|^2\cr
\big\| |\De+1;\ell-1 \r \big\|^2 & {} =
16(\De-\ell-1)\ell^2(4\ell^2-1)\big\| |\De;\ell\r \big\|^2\,.}
}
Another important example is for $SO(4,2)$ whereby
along with \coseventeen, for
$\uell=(\ell_1,\ell_2)$, we also have
\eqn\coseventeen{\eqalign{
& |\De+1;\ell_1,\ell_2-1\r=
(-2i(\ell_1+\ell_2)\P_{2-}
+\P_{1+} E_{12}^{--}) |\De;{\uell}\r^{\rhw} \,,\cr
& |\De+1;\ell_1-1,\ell_2\r=
\big(-4(\ell_1^2-\ell_2^2)\P_{1-}-2i(\ell_1-\ell_2)\P_{2+}E_{12}^{--}\cr
& \qquad \qquad \qquad \qquad \qquad
-2i(\ell_1+\ell_2)\P_{2-}E_{12}^{-+}+\P_{1+}
E_{12}^{-+}E_{12}^{--}\big)
|\De;{\uell}\r^{\rhw} \,,
}}
for which the norms are
\eqn\threestates{\eqalign{
\big\| |\De+1;\ell_1,\ell_2-1\r \big\|^2 & {} =
16 (\De-\ell_2-1)(\ell_1+\ell_2)(\ell_1+\ell_2+1)
\big\| |\De;\uell\r \big\|^2\,,\cr
\big\| |\De+1;\ell_1-1,\ell_2\r \big\|^2 & {} =
64 (\De-\ell_1-2)(\ell_1^2-\ell_2^2)(\ell_1-\ell_1+1)(\ell_1+\ell_2+1)
\big\| |\De;\uell\r \big\|^2\,.}
}

Constructing other such elements of $\H_d$ is cumbersome.
We here outline a simpler procedure
for finding the unitarity constraints  for $\B^{(1)}$.
The norms of the highest weight states
in $\H_d$ are more generally given by,
\eqn\norms{
\big\| |\De+1;\uell+\uv
\r\big\|^2=\Big(\De+g^{(\v)}_{\uell}\Big)f^{(\v)}_{\uell}\,,
}
where the functions $f^{(\v)}_{\uell}$ are strictly positive for
$\ell$ being strictly inside the
dominant Weyl chamber, \piffle\ or \piffles.  We have that, 
assuming that $\ell$ is strictly inside the dominant Weyl
chamber,
\eqn\whatwehavefor{
\K_{1-}|\De+1;\uell+\uv\r=0\quad \Rightarrow \quad \De+g^{(\v)}_{\uell}=0\,,
}
which is in turn implied by the state $|\De+1;\uell+\uv\r$ being
null.
As an aid to solving \whatwehavefor\ we
extend the definition of $\B^{(1)}$ in \copo\ and consider $\ell'\in \V_\ell$
(the Verma module with dominant integral highest weight $\ell$).
 Consider the following highest weight 
states with respect to
$SO(d)$, namely,
\eqn\tryagain{
|\De+1;\uell^{w_\v}+\ue_1\r=\P_{1+}|\De;\uell^{w_\v}\r\,,\quad \K_{1-}
|\De+1;\uell^{w_\v}+\ue_1\r=0\quad \Rightarrow\quad \De+\ell_1^{w_{\v}}=0\,,
}
where $w_\v$ are such members of the Weyl group $\W_r$ for which
\eqn\prest{
\ell^{w_{\v}}+\e_1=(\ell+w_{\v}{}^{-1}(\e_1))^{w_\v}=(\ell+\v)^{w_\v}\,,
}
for some $\v=\vep \e_j$ for $\vep=\pm$.{\foot{Note that for $SO(2r+1)$
the
vector $\v=0$ is not on the Weyl orbit of $\e_1$ - we need a different
approach to deal with this.}}
Thus the states \tryagain\ are related to 
those in $\H_d$
by the action of $SO(d)$ lowering operators on
$|\De+1;\uell+\uv\r$.{\foot{ 
 For instance, we have that  
$|\De+1;\uell^{\sigma_{12}}+{\underline{e}}_1\rangle
=(E_{12}^{-+})^{\ell_1-\ell_2}|\De+1;\uell+{\underline{e}}_2\rangle$
 for $\sigma_{12}(\ell_1,\ell_2,\dots)=(\ell_2,\ell_1,\dots)$
and with $|\De{+1},\uell+\ue_2\rangle$ given in
\coseventeen.}}
Also, $\K_{1-}$ commutes with all such
lowering operators so that the conditions \whatwehavefor\
and \tryagain\ should be identical. Thus, 
 using \compts\ (for $\vep_1=\vep,\,\si(1)=j$),
\eqn\ribbetsev{
g^{(\vep\e_j)}_{\uell}=\ell_1^{w_{\vep\e_j}}=
\vep\ell_j+(\vep-1)\half d -\vep j+1\,,
}
determining $g^{(\v)}_{\uell},\,\v\neq 0$ in \norms\ for the
states in $\H_d$.

To deal with the state $|\De+1;\ell\r \in \H_{2r+1}$ for $SO(2r+1,2)$
we first note an interesting observation.  Consider the
state $|\De+1;\uell+\ue_{r+1}\r\in \H_{2r+2}$ which is given by
\eqn\statedef{\eqalign{
&|\De+1;\uell+\ue_{r+1}\r\cr
&{}\,\,=
\Big(A_{\uell}\,\P_{r+1\,+}+\!\!\!\sum_{1\leq i\leq r,\,\si}\!\!\!
 B_{\uell,i,\si}\,\P_{i+}E_{\si(i\,i+1)}^{-+}
E_{\si(i+1\,i+2)}^{-+}\dots 
E_{\si(r\,r+1)}^{-+}\Big)|\De,\uell\r^{\rhw}\,,}
}
where $\si$ permutes $(i\,i+1),\dots, (r\, r+1)$
and $A_{\uell},\,B_{\uell,i,\si}$ are determined from the
requirement that $E_{12}^{+-},\dots E_{r\,r+1}^{+-}$ annihilate
the state ($E_{r\,r+1}^{++}$ automatically annihilates it).
Defining ${\tilde A}_{(\ell_1,\dots,\ell_r)}=A_{(\ell_1,\dots,
\ell_r,0)},\,{\tilde B}_{(\ell_1,\dots,\ell_r),i,\si}=
B_{(\ell_1,\dots,\ell_r,0),i,\si}$, then we claim that
$|\De+1;\ell\r \in \H_{2r+1}$ is given by
\eqn\statedefo{\eqalign{
&|\De+1;\uell\r=
\Big({\tilde A}_{\uell}\,\P_{0}+\!\!\!\sum_{1\leq i\leq r,\,\si}\!\!\!
 {\tilde B}_{\uell,i,\si}\,\P_{i}^{+}E_{\si(i\,i+1)}^{-+}
E_{\si(i+1\,i+2)}^{-+}\dots 
E_{\si(r)}^{-}\Big)|\De,\uell\r^{\rhw}\,,}
}
where now $\si$ permutes $(i\,i+1),\dots (r)$.
This follows when we show that the conditions on 
${\tilde A}_{\uell},\tilde{B}_{\uell,i,\si}$
arising from
$E_{i\,i+1}^{+-},\,E_{r}^{+},\,1\leq i\leq r-1$ annihilating \statedefo\ are
exactly equivalent to those on $
A_{\uell},B_{\uell,i,\si}$
arising from $E_{i\,i+1}^{+-},\,
E_{r\,r+1}^{+-},\,1\leq i\leq r-1$ annihilating \statedef\ for
$\ell_{r+1}=0$ if we identify $\P_{0}$ with $\P_{r+1}$ and
$E_{r}^{-}$ with $E_{r\,r+1}^{-+}$.  
We have that $[\K_{1-},\P_{r+1\,+}]=-2i E_{1\,r+1}^{ -+}=(-2i)^{1-r}
[E_{12}^{-+},
\dots ,[E_{r-1\,r}^{-+},E_{r\,r+1}^{-+}]\dots ]$ and
$[\K_{1-},\P_0]=-2i E_{1}^{-}=(-2i)^{1-r}[E_{12}^{-+},
\dots ,[E_{r-1\,r}^{-+},E_{r}^{-}]\dots ]$.   Due to
this and as
${\tilde A}_{(\ell_1,\dots,\ell_r)}=A_{(\ell_1,\dots,
\ell_r,0)},\,{\tilde B}_{(\ell_1,\dots,\ell_r),i,\si}=
B_{(\ell_1,\dots,\ell_r,0),i,\si}$ then $\K_{1-}$ annihilating
\statedef\ for $\ell_{r+1}=0$ results in the same equations for
$\De$ as for $\K_{1-}$ annihilating $\statedefo$ if we identify
$E_{r\,r+1}^{-+}$ with $E_{r}^{-}$.  
  Thus, from \ribbetsev\ for $j=r+1,\,\vep=+,\,\ell_{r+1}=0$,
\eqn\frump{
\K_{1-}|\De+1,\uell\r=0\quad \Rightarrow \quad g^{(0)}_{\uell}=-r\,.
}

Now that we have determined $g^{(\v)}_{\uell}$ in \norms\ 
to be given by \ribbetsev\ and \frump, we may
determine the unitarity bounds for states in $\B^{(1)}$,
the simplest descendants.
For $\H_{d}$ and
$\ell_1=\dots=\ell_p>|\ell_{p+1}|,\,p\leq r-1$ then we have that
$f^{(\v)}_{\uell}=0$ in \norms\ 
for $\v=-\e_1,\vep \e_j,\e_p,\,j=2,\dots,p-1${\foot{Here and in the
following,
this is because the corresponding states $|\De+1;\uell+\uv \rangle$
for $\ell$ being on the boundary of the dominant Weyl chamber
are null as $SO(d)$ representations.}}
and that
\eqn\unitarityboundone{\eqalign{
\De&{} \geq {\rm
max.}\{-g_{\uell}^{(\e_1)}
,-g_{\uell}^{(-\e_p)}
,-g_{\uell}^{(\e_j)},-g^{(-\e_j)}_{\uell},\,\,\,p+1\leq j\leq r\}\cup
\{-g^{(0)}_{\uell}\,\,{\rm for}\,\, d=2r+1\}\cr
&{} =-g_{\uell}^{(-\e_p)}=\ell_1+d-p-1\,,}
} 
which matches the first requirement in \unitarityd.
At the unitarity bound $\De=\ell_1+d-p-1$ then all the states
$|\De+1;\uell+\uv\r$ for $\v=-\e_1,\vep \e_j,\,2\leq j\leq p$ in
$\H_{d}$ 
are
null. 

In even dimensions, for $\ell_1=\dots=\pm\ell_r$ for
$\H_{2r}$
then we have that $f^{(\v)}_{\uell}=0$
 in \norms\ 
for $\v=-\e_1,\vep \e_j,\pm\e_r,\,j=2,\dots,r-1$ 
\eqn\unitarityboundtwo{
\De\geq {\rm max.}\{-g_{\uell}^{(\e_1)},-g_{\uell}^{(\mp\e_r)}\}=\ell_1+r-1\,,
}
 with in addition the state 
$|\De+1;\uell\mp\ue_r\r$ being null at the unitarity bound. 

In odd dimensions, for $\ell_1=\dots =\ell_r> {1\over 2}$
for $\H_{2r+1}$ then $f^{(\v)}_{\uell}=0$ in \norms\ 
for $\v=-\e_1,\vep \e_j,\e_r,\,j=2,\dots,r-1$ and 
\eqn\unitarityboundthree{
\De\geq {\rm
max.}\{-g_{\uell}^{(\e_1)},\,
-g_{\uell}^{(-\e_r)},\,-g^{(0)}_{\uell}\}=\ell_1+r\,,
}
with in addition the state
$|\De+1,\uell-\ue_r\r$
being null at the unitarity bound.  
For $\ell_1=\dots =\ell_r=\half$
then $f^{(\v)}_{\uell}=0$ in \norms\ 
for $\v=-\e_1,\vep \e_j,\,j=2,\dots,r$ and 
\eqn\unitarityboundfour{
\De\geq {\rm max}.\{-g_{\uell}^{(\e_1)},\,-g^{(0)}_{\uell}\}=r=[\half d]\,,
} 
with the
state $|\De+1,\uell\r$ being null at the unitarity bound.

\appendix{D}{Expansion and product formulae}

In this appendix various formulae from section four are
proven. We make use of a simple property of the function 
$P^{(d)}(s,\x)$ defined for $d=2r$ in \defP\ and $d=2r+1$ in \defPo.
Under the action of the Weyl symmetry operator $\frak{W}_d$
(defined in appendices A,B) it obeys
\eqn\actionPe{
{\frak W}_{d}\big(f(s,\x)P^{(d)}(s,\x)\big)=
{\frak W}_{d}\big(f(s,\x)\big)\,\,P^{(d)}(s,\x)\,,
}
for any $f(s,\x)$, as $P^{(d)}(s,\x)$ is invariant under the action
of any element of the $SO(d)$ Weyl group, $\W_{d}$.
Note also that $\W_{d}$ has no effect on the variable $s$.

We discuss the even dimensional cases of
 \findthatb, \csixteenanal, \csixteenanalopo\ first.
For \csixteenanal\ we have that,
\eqn\firstproof{\eqalign{
\sum_{q=0}^{\infty}s^{\ell+r+q-1} 
\chi^{(2r)}_{(\ell+q,\ell,\dots,\pm\ell)}(\x)&{}=
s^{\ell+r-1}{\frak W}_{2r}\Big(\sum_{q=0}^{\infty}(s\,x_{1})^q 
C^{(2r)}_{(\ell,\dots,\pm\ell)}(\x)\Big)\cr
&{}=s^{\ell+r-1}{\frak W}_{2r}\Big({1\over 1-s
x_1}C^{(2r)}_{(\ell,\dots,\pm\ell)}(\x)\Big)\,,}
}
which follows just by the definition of
the character of the irreducible representation \coatwo\ in
terms of the Verma module character \coazero.
Using \actionPe\ then \firstproof\ may be
rewritten as
\eqn\secondproof{\eqalign{
& s^{\ell+r-1}P^{(2r)}(s,\x){\frak W}_{2r}\Big((1-s x_1{}^{-1})
\prod_{i=2}^{r}(1-s x_i)(1-s
x_i{}^{-1})C^{(2r)}_{(\ell,\dots,\pm\ell)}(\x)\Big)\cr
&{}=s^{\ell+r-1}P^{(2r)}(s,\x)\!\!\!\!\!\!\!\!
\sum_{{n_{i-},n_{j+}=0,1
\atop 0\leq n=\sum n_{i\vep}\leq 2r-2}}\!\!\!\!\!\!\!\!
(-s)^n \chi^{(2r)}_{(\ell-n_{1-},\ell+n_{2+}-n_{2-},\dots,
\pm\ell+n_{r+}-n_{r-})}(\x)\,.
}}
For $n>0$ in \secondproof\ we may use 
\eqn\formula{
\chi^{(d)}_{(\ell_1,\dots,\ell_j,\ell-1,\ell+1,\ell_{j+3},\dots,\ell_r)}(\x)=
-\chi^{(d)}_{(\ell_1,\dots,\ell_j,\ell,\ell,\ell_{j+3},\dots, \ell_r)}(\x)\,,}
to show that the contributions for given $n$
reduce to a single one from
\eqn\contrib{
\chi^{(2r)}_{{(\ell,\dots,\ell,\ell-1,\dots,\ell-1,\pm\ell\mp 1)\atop{
\!\!\!\!\!\!\!\!\!\!\!\!\!
\uparrow\atop n^{\rm th}\,\,{\rm position}}}}(\x)\,,
}
with all contributions for $n>r$ vanishing.
Hence we have that \firstproof\ reduces to
$\D^{(2r)}_{[\ell+r-1,\ell]\pm}(s,\x)$
defined in \defs\ thus proving \csixteenanal.  
Notice that \findthatb\ is a special case of \csixteenanal\ 
when we take $\ell=0$ in the latter (whereby, as mentioned before,
$\D^{(2r)}_{[\ell+r-1,\ell]\pm}(s,\x)\to s^{r-1}(1-s^2)P^{(2r)}(s,\x)$).

We may prove \csixteenanalopo\ in a very similar way.
The sum on the right hand side  of \csixteenanalopo\ 
may be reduced to
\eqn\reductionpoint{
s^{\ell+r+j-1}
{\frak W}_{2r}\Big((1-s x_1)^{-1}\prod_{i={r-j+1}}^{r}(1-s x_i)^{-1}(1-s
x_i{}^{-1})^{-1}\,
C^{(2r)}_{(\ell,\dots,\ell,\ell_1,\dots,\ell_j)}(\x)\Big)\,,
}
in a similar fashion as \firstproof, when we perform the sums
over $p_1,\dots,p_j,q$.
This may be rewritten using \actionPe\ as
\eqn\reductcontd{\eqalign{
&{}s^{\ell+r+j-1}P^{(2r)}(s,\x){\frak W}_{2r}\Big((1-s x_1)\prod_{i=2}^{r-j}
(1-s x_i)(1-s
x_i{}^{-1})
C^{(2r)}_{(\ell,\dots,\ell,\ell_1,\dots,\ell_j)}(\x)\Big)\cr
&{}=s^{\ell+r+j-1}P^{(2r)}(s,\x)\!\!\!\!\!\!\!\!\sum_{{n_{i-},n_{j+}=0,1
\atop 0\leq n=\sum n_{i\vep}\leq 2r-2j-2}}\!\!\!\!\!\!\!\!
(-s)^n \chi^{(2r)}_{(\ell-n_{1-},\ell+n_{2+}-n_{2-},\dots,
\ell+n_{r{-j}\,+}-n_{r{-j}\,-},\ell_1,\dots,\ell_r)}(\x)\,.}
}
For similar reasons as before, for $n>0$ the contributions for given $n$
reduce to a single one from
\eqn\contrib{
\chi^{(2r)}_{{(\ell,\dots,\ell,\ell-1,\dots,\ell-1,\ell_1,\dots, \ell_j)\atop{
\!\!\!\!\!\!\!\!\!\!\!\!\!\!\!\!\!
\uparrow\atop n^{\rm th}\,\,{\rm position}}}}(\x)\,,
}
so that \reductcontd\ equals 
$\D^{(2r)}_{[\ell+r+j-1,\ell,\ell_1,\dots,\ell_j]}(s,\x)$ in
\defr\ for $p=r-j$.

Turning to the odd dimensional cases of \findthatbk, \expoddo,
\expoddoro, \csixteenanalopodd\ these may be proven in a very similar
way
as for the even dimensional cases when we use \actionPe.
Note that we may use the definition of the
irreducible character 
\chaross\ in terms of the Verma module character
\coazerop\ and \defPo\ to rewrite the sum on the right
hand side of \expoddoro\ as
\eqn\simplifexpodd{\eqalign{
& s^r P^{(2r+1)}(s,\x)(1-s){\frak
W}_{2r+1}\Big((1-s x_1{}^{-1})\prod_{i=2}^{r}
(1-s x_i)(1-s x_i{}^{-1})\,C^{(2r+1)}_{({1\over 2},\dots,{1\over
2})}(\x)\Big)\cr
& = s^r P^{(2r+1)}(s,\x)(1-s)\chi^{(2r+1)}_{({1\over 2},\dots,{1\over
2})}(\x)\,,
}}
which matches  \defto.
 The free scalar case of  \expoddo\
follows in a similar fashion.  The identity \findthatbk\ is in
fact equivalent to \expoddo.
The sum on the right hand side of \csixteenanalopodd\
may be rewritten as
\eqn\reductionpointodd{
s^{\ell+r+j}
{\frak W}_{2r+1}\Big((1-s)^{-1}
(1-s x_1)^{-1}\prod_{i={r-j+1}}^{r}(1-s x_i)^{-1}(1-s
x_i{}^{-1})^{-1}\,
C^{(2r+1)}_{(\ell,\dots,\ell,\ell_1,\dots,\ell_j)}(\x)\Big)\,,
}
when we perform the sums over $p_i,q,t$.
This may be rewritten as
\eqn\reductcontdodd{\eqalign{
&{}s^{\ell+r+j}P^{(2r+1)}(s,\x){\frak W}_{2r+1}\Big((1-s x_1)\prod_{i=2}^{r-j}
(1-s x_i)(1-s
x_i{}^{-1})
C^{(2r+1)}_{(\ell,\dots,\ell,\ell_1,\dots,\ell_j)}(\x)\Big)\cr
&{}=s^{\ell+r+j}P^{(2r+1)}(s,\x)\!\!\!\!\!\!\!\!\sum_{{n_{i-},n_{j+}=0,1
\atop 0\leq n=\sum n_{i\vep}\leq 2r-2j-2}}\!\!\!\!\!\!\!\!
(-s)^n \chi^{(2r+1)}_{(\ell-n_{1-},\ell+n_{2+}-n_{2-},\dots,
\ell+n_{r{-j}\,+}-n_{r{-j}\,-},\ell_1,\dots,\ell_r)}(\x)\,.}
}
Using \formula, for $n>0$ the contributions for given $n$
reduce to a single one from
\eqn\contrib{
\chi^{(2r+1)}_{{(\ell,\dots,\ell,\ell-1,
\dots,\ell-1,\ell_1,\dots, \ell_j)\atop{
\!\!\!\!\!\!\!\!\!\!\!\!\!\!\!\!\!
\uparrow\atop n^{\rm th}\,\,{\rm position}}}}(\x)\,,
}
so that \reductcontdodd\ equals 
$\D^{(2r+1)}_{[\ell+r+j,\ell,\ell_1,\dots,\ell_j]}(s,\x)$ in
\defro\ for $p=r-j$.

To prove some product formulae we will use the
expansions above and the following,
\eqn\charinv{
{\frak W}_d\big(f(\x)\chi^{(d)}_{\uell}(\x)\big)=
{\frak W}_d\big(f(\x))\,\chi^{(d)}_{\uell}(\x)\,,
}
for any $f(\x)$.

For even dimensions we now prove \productonetp\
and
\productonet\ for $\ell'={1\over 2}$.
In this case we have that
\eqn\prodhalfo{
\D^{(4m+2)}_{[{1\over 2}+2m;{1\over 2}]\pm}(s,\x)=s^{{1\over 2}+2m}
\big(\chi^{(4m+2)}_{({1\over 2},\dots,{1\over 2},\pm{1\over 2})}(\x)
-s \chi^{(4m+2)}_{({1\over 2},\dots,{1\over 2},\mp{1\over
 2})}(\x)\big)\,,
}
where we may determine from \coatwo\ that,
\eqn\chihalfevp{\eqalign{
\chi^{(4m+2)}_{({1\over 2},\dots,{1\over 2})}(\x)
&{} =\prod_{i=1}^{2m+1} x_i{}^{-{1\over 2}}\!\!\!\!\!\!\!\!\!\!\!\!\!\!
\sum_{{m\geq t\geq 0\atop 2m+1\geq j_1>\dots>j_{2t+1}\geq 1}}
\!\!\!\!\!\!\!\!\!\!\!\!\!\!x_{j_1}\cdots x_{j_{2t+1}}\,,\cr
\chi^{(4m+2)}_{({1\over 2},\dots,-{1\over 2})}(\x)
&{} =\prod_{i=1}^{2m+1} x_i{}^{-{1\over 2}}\Big(1+\!\!\!\!\!\!\!\!\!\!\!\!\!\!
\sum_{{m\geq t\geq 1\atop 2m+1\geq j_1>\dots>j_{2t}\geq 1}}
\!\!\!\!\!\!\!\!\!\!\!\!\!\!x_{j_1}\cdots x_{j_{2t}}\Big)\,.}
}

We use \prodhalfo\ and \csixteenanal\ 
to expand $\D^{(4m+2)}_{[\ell+2m,\ell]\pm}(s,\x)$ in \productonetp\
and then match powers of $s$ on both
sides.  Clearly the $O(1)$ terms on both sides of \productonetp\ agree.
At $O(s^q)$ for $q\geq 1$ we must show that,
\eqn\whatweneedtsp{\eqalign{
\!\!\!\!& \chi^{(4m+2)}_{(\ell+q,\ell,\dots,\ell)}(\x)
\chi^{(4m+2)}_{({1\over 2},\dots,-{1\over 2})}(\x)
- \chi^{(4m+2)}_{(\ell+q-1,\ell,\dots,\ell)}(\x)
\chi^{(4m+2)}_{({1\over 2},\dots,{1\over 2})}(\x)\cr
&{} =
\sum_{t_i=\ell\pm{1\over 2}\atop t_i\geq t_{i+1}}
\chi^{(4m+2)}_{(\ell+{1\over 2}+q,\ell+{1\over 2},
t_1,t_1,\dots,t_{m-1},t_{m-1},\ell-{1\over 2})}(\x)
-\chi^{(4m+2)}_{(\ell+{1\over 2}+q-2,\ell+{1\over 2},
t_1,t_1,\dots,t_{m-1},t_{m-1},\ell-{1\over 2})}(\x)\,.}
}
Using \charinv\ and \chihalfevp\ 
we may rewrite the left hand side of \whatweneedtsp\ as
\eqn\rhswhatweneedtsp{\eqalign{
& {\frak W}_{4m+2}\big(C^{(4m+2)}_{(\ell+q,\ell,\dots,\ell)}(\x)
\chi^{(4m+2)}_{({1\over 2},\dots,-{1\over 2})}(\x)
- C^{(4m+2)}_{(\ell+q-1,\ell,\dots,\ell)}(\x)
\chi^{(4m+2)}_{({1\over 2},\dots,{1\over 2})}(\x)\big)\cr
& ={\frak W}_{4m+2}\bigg(
C^{(4m+2)}_{(\ell+q-{3\over 2},\ell-{1\over 2},\dots,\ell-{1\over
2})}(\x)
(x_{1}{}^2-1)\!\!\!\!\!\!\!\!\!\!\!\!\!\!
\sum_{{m\geq t\geq 1\atop 2m+1\geq j_1>\dots>j_{2t-1}\geq 2}}
\!\!\!\!\!\!\!\!\!\!\!\!\!\!x_{j_1}\cdots x_{j_{2t-1}}\bigg)\,.
}}
For $q\geq 1$ most of the terms in \rhswhatweneedtsp\ vanish 
under the action of the Weyl symmetry operator and it
reduces to
\eqn\reductiop{
 {\frak W}_{4m+2}\bigg(
C^{(4m+2)}_{(\ell+q-{3\over 2},\ell-{1\over 2},\dots,\ell-{1\over
2})}(\x)
(x_{1}{}^2-1)\sum_{t= 1}^m
x_{2}x_{3}\cdots x_{2t}\bigg)\,,
}
and from here it is easy to show that this agrees with the
right hand side of
\whatweneedtsp.

Similarly, using \prodhalfo\ and \csixteenanal\ 
to expand $\D^{(4m+2)}_{[\ell+2m,\ell]\pm}(s,\x)$ in \productonet,
then matching powers of $s$ on both
sides of the equation \productonet\ we must show that for $q\geq 0$,
\eqn\whatweneedts{\eqalign{
& \chi^{(4m+2)}_{(\ell+q,\ell,\dots,\vep \ell)}(\x)
\chi^{(4m+2)}_{({1\over 2},\dots,\vep{1\over 2})}(\x)
- \chi^{(4m+2)}_{(\ell+q-1,\ell,\dots,\vep \ell)}(\x)
\chi^{(4m+2)}_{({1\over 2},\dots,-\vep{1\over 2})}(\x)\cr
&{} =
\sum_{t_i=\ell\pm{1\over 2}\atop t_i\geq t_{i+1}}
\chi^{(4m+2)}_{(\ell+{1\over 2}+q,t_1,t_1,\dots,t_{m},\vep t_m)}(\x)
-\chi^{(4m+2)}_{(\ell+{1\over 2}+q-2,t_1,t_1,\dots,t_{m},\vep
 t_m)}(\x)\,,}
}
for $\vep=\pm$.
Using \charinv\ and \chihalfevp\ 
we may rewrite the left hand side of \whatweneedts\ for $\vep=+$ as
\eqn\rhswhatweneedts{\eqalign{
& {\frak W}_{4m+2}\big(C^{(4m+2)}_{(\ell+q,\ell,\dots,\ell)}(\x)
\chi^{(4m+2)}_{({1\over 2},\dots,{1\over 2})}(\x)
- C^{(4m+2)}_{(\ell+q-1,\ell,\dots,\ell)}(\x)
\chi^{(4m+2)}_{({1\over 2},\dots,-{1\over 2})}(\x)\big)\cr
& ={\frak W}_{4m+2}\bigg(
C^{(4m+2)}_{(\ell+q-{3\over 2},\ell-{1\over 2},\dots,\ell-{1\over
2})}(\x)
(x_{1}{}^2-1)\Big(1+\!\!\!\!\!\!\!\!\!\!\!\!\!\!
\sum_{{m\geq t\geq 1\atop 2m+1\geq j_1>\dots>j_{2t}\geq 2}}
\!\!\!\!\!\!\!\!\!\!\!\!\!\!x_{j_1}\cdots x_{j_{2t}}\Big)\bigg)\,.
}}
For $q\geq 0$ most of the terms in \rhswhatweneedts\ vanish 
under the action of the Weyl symmetry operator and it
reduces to
\eqn\reductio{
 {\frak W}_{4m+2}\bigg(
C^{(4m+2)}_{(\ell+q-{3\over 2},\ell-{1\over 2},\dots,\ell-{1\over
2})}(\x)
(x_{1}{}^2-1)\Big(1+
\sum_{t= 1}^m
x_{2}x_{3}\cdots x_{2t+1}\Big)\bigg)\,,
}
and from here it is easy to show that this agrees with the
right hand side of
\whatweneedts\ for $\vep=+$.

We also have that
\eqn\prodhalfo{
\D^{(4m)}_{[{1\over 2}+2m-1;{1\over 2}]\pm}(s,\x)=s^{{1\over 2}+2m-1}
\big(\chi^{(4m)}_{({1\over 2},\dots,{1\over 2},\pm{1\over 2})}(\x)
-s \chi^{(4m)}_{({1\over 2},\dots,{1\over 2},\mp{1\over
 2})}(\x)\big)\,,
}
where,
\eqn\chihalfevp{\eqalign{
\chi^{(4m)}_{({1\over 2},\dots,-{1\over 2})}(\x)
&{} =\prod_{i=1}^{2m} x_i{}^{-{1\over 2}}\!\!\!\!\!\!\!\!\!\!\!\!\!\!
\sum_{{m-1\geq t\geq 0\atop 2m\geq j_1>\dots>j_{2t+1}\geq 1}}
\!\!\!\!\!\!\!\!\!\!\!\!\!\!x_{j_1}\cdots x_{j_{2t+1}}\,,\cr
\chi^{(4m)}_{({1\over 2},\dots,{1\over 2})}(\x)
&{} =\prod_{i=1}^{2m} x_i{}^{-{1\over 2}}\Big(1+\!\!\!\!\!\!\!\!\!\!\!\!\!\!
\sum_{{m\geq t\geq 1\atop 2m\geq j_1>\dots>j_{2t}\geq 1}}
\!\!\!\!\!\!\!\!\!\!\!\!\!\!x_{j_1}\cdots x_{j_{2t}}\Big)\,,}
}
and this allows similar product formulae in
$d=4m$ dimensions to be derived 
straightforwardly in an analogous fashion as above.

\listrefs

\bye